\documentclass{article}
\usepackage{graphicx}
\usepackage{subfigure}

\begin{document}

\baselineskip=23pt

\begin{center}

\textbf{\Huge The Neighborhood Function and Its Application to
Identifying Large-Scale Structure in the Comoving Universe
Frame}\footnote{Send offprint requests to: Y.-P. Qin
(ypqin\_cn@yahoo.com.cn), L.-Z. L\"{u}(llz@ynao.ac.cn)}

\vspace{2mm}

\textbf{Y.-P. Qin$^{1,2,3}$, L.-Z. L\"{u}$^{2,3,4}$, F.-W.
Zhang$^{2,4}$, B.-B. Zhang$^{2,4}$ and J. Zhang$^{2,4}$}

\textbf{$^1$Center for Astrophysics, Guangzhou University, Guangzhou
510006, P. R. China}

\textbf{$^2$National Astronomical Observatories/Yunnan Observatory,
Chinese Academy of Sciences, Kunming 650011, P. R. China}

\textbf{$^3$Physics Department, Guangxi University, Nanning 530004,
P. R. China}

\textbf{$^4$ The Graduate School of the Chinese Academy of Sciences,
Beijing 100049, P. R. China}

\end{center}

\vspace{4mm}

\begin{abstract}
In this paper, a relative number density parameter, called the
neighborhood function, is introduced so that the crowded nature of
the neighborhood of individual sources could be described. With this
parameter one can determine the probability of forming a cluster
with a given number of galaxy members and a certain number density
by chance. A method is proposed to identify large-scale structure on
the cosmological co-moving frame. To avoid those effects arising
from the distance of objects when applying the method, we compare
the real number density concerned with the mean number density
measured in the corresponding local area rather than the mean of the
whole sample. The scale used to sort out clustering sources is
determined by that mean local number density, and thus it is a
redshift-dependent scale. In applying this method, the sample
adopted is required to have regular borders so that a simulation
over the sample area could be performed and the volume confined by
the sorting out scale within the sample area could be determined and
then the probability referring to number density could be evaluated.
The method is applied to a sample drawn from the 2dF survey,
analyzed in redshift space. We find from the analysis that the
probability of forming the resulting large-scale structures by
chance is very small and the phenomenon of clustering is dominant in
the local Universe. Within the $3\sigma $ confidence level, a
coherent cluster with its scale as large as $357h^{-1}Mpc$ and
another with its number of galaxy members as large as 12966 are
identified from the sample. There exist some galaxies which are not
affected by the gravitation of clusters and hence are suspected to
rest on the co-moving frame of the Universe. Voids are likely the
volumes within which no very crowded sources are present and they
are likely formed in embryo by fluctuation in the very early epoch
of the Universe. In addition, we find that large-scale structures
are coral-like and they are likely made up of smaller ones; sources
with large values of the neighborhood function are mainly
distributed within the structure of prominent clusters and it is
them who form the frame of the large-scale structure.
\end{abstract}

\begin{flushleft}
\end{flushleft}

\vspace{4mm}

\section{Introduction}\label{sec.intro}

Distributions of galaxies in the sky were found useful in studying
large-scale structures of the Universe (Seldner et al. 1977; Gellar
\& Huchra 1989; Loveday et al. 1992). Owing to redshift surveys
available in different eras, many investigations were made to detect
large-scale structures in both the projected map and the
three-dimensional space. There were many discoveries of clustering
features in 80's, including voids, galaxy clusters, strings, great
attractors, and various great walls (Kirshner et al. 1981; Bahcall
\& Soneira 1983; Haynes \& Giovarelli 1986; Lynden-Bell et al. 1988;
Gellar \& Huchra 1989). In as early as 1989, the CfA ``Great Wall''
was found to extend to $170 h^{-1} Mpc$ (Gellar \& Huchra
1989)\footnote{No particular cosmological models were adopted in
Gellar \& Huchra (1989) to calculate the length of the structure due
to the fact that the adopted redshifts are small.}, while in recent
years the Sloan Great Wall was observed to be stretched to $\sim 400
h^{-1} Mpc$ in length by Gott III et al. (2005)\footnote{In Gott III
et al. (2005), the adopted cosmological parameter is
$\Omega_{M}=0.27$.} and by Deng et al. (2006)\footnote{In their
analysis, Deng et al. (2006) adopted $\Omega_{M}=0.3$. Hereafter in
this section, when a cosmological model is adopted in a cited paper,
the parameter will be $\Omega_{M}=0.3$ unless otherwise specified.}
respectively. The latter is the largest structure of the Universe
observed so far. Observations also revealed that large-scale
structures could be found deep in the space where redshift is high
(Shimasaku et al. 2003; Matsuda et al. 2005; Ouchi et al. 2005).
Investigation of large-scale structures of the Universe meets a new
chance owing to two recent redshift surveys, the 2-degree Field
galaxy redshift survey (2dFGRS) ( Colless et al. 2001) and the Sloan
Digital Sky Survey (SDSS) (York et al. 2000) which provide large
amount and reliable data.

Large-scale structures of the local Universe are of particular
importance since theories of structure formation could be directly
checked by redshift surveys. It was predicted that large-scale
filamentary or sheet-like mass overdense regions would
preferentially be formed in the early Universe and later evolve into
dense clusters of galaxies which would be expected in the local
Universe, and gravitational amplification of the matter density
fluctuations that are generated in the early Universe is assumed to
be responsible for the formation of the clustering structure of the
present Universe (see Peebles 1982; Blumenthal et al. 1984; Davis et
al. 1985, 1992; Governato et al. 1998; Kauffmann et al. 1999; Cen \&
Ostriker 2000; Benson et al. 2001; Colberg et al. 2005). Filamentary
features were found to be real when some redshift surveys were
carefully checked (Bhavsar \& Ling 1988; Bharadwaj et al. 2000;
Bagchi et al. 2002; Ebeling et al. 2004; Pimbblet et al. 2004;
Porter \& Raychaudhury 2005). The detection of a large concentration
of primeval galaxies at redshift $z\sim 3$ (Steidel et al. 1998) and
the reported discovery of a large-scale coherent filamentary
structure of Ly$ \alpha $ emitters in a redshift space at $z=3.1$
favor the prediction that galaxies preferentially formed in
large-scale filamentary or sheet-like mass overdensities in the
early Universe (Matsuda et al. 2005). In agreement with these,
protoclusters which have been presumed not to have virialized due to
their short cosmological ages (e.g. Venemans 2005) were identified
at as far as $z\sim 4$ (Venemans et al. 2002; Intema et al. 2006).

The great success of cold dark matter (CDM) model in recent years
makes it a leading model in current studies of cosmology. The charm
of the CDM model is its predictive power on the formation of
large-scale structures. However, in this model, large-scale
structures up to $>100h^{-1} Mpc$\ are rarely formed, which are
challenged by recent observations (see Yoshida et al. 2001; Yoshida
2005).  It suggests that the CDM model predicts smaller homogeneity
scale of the Universe than what the observations have revealed (note
that the large-scale structure of the Universe was found to be
several hundred $h^{-1} Mpc$). In addition, the model faces
difficulty in explaining the observed structure on length scales
$<1h^{-1} Mpc$ as well (see a brief review in Cembranos et al.
2005).

Statistical methods are useful in studying large-scale structures of
the Universe. As mentioned in Martinez \& Saar (2002a), most of the
statistical analysis of the galaxy distribution proposed so far are
based on second order methods (correlation functions and power
spectra) (see also Diggle 1983). Among them, the two-point
correlation function (Totsuji \& Kihara 1969; Peebles 1974, 1980;
Davis \& Peebles 1983) and the power spectrum (Fisher et al. 1993;
Feldman et al. 1994; Park et al. 1994; Tadros \& Efstathiou 1996)
are the tools most often used in both observational and theoretical
analyses. The latter is the Fourier transform of the former. Since
the two quantities are a Fourier transform pair, complete knowledge
of one is equivalent to complete knowledge of the other. But this is
not true for their estimators when samples employed are finite and
noisy (Feldman et al. 1994). The two-point correlation function $\xi
(r)$, which was first used to measure the strength of galaxy
clustering for redshift surveys by Davis \& Peebles (1983),
describes the excess probability of finding a galaxy in a volume
element at a separation $r$ from another randomly chosen galaxy
above that for an un-clustered distribution. There are various
estimators of $\xi (r)$ in literature. The main difference is their
corrections for the edge effect (Martinez \& Saar 2002a). The
quantity was found to obey a power law: $\xi (r)=(r/r_0)^{-\gamma }$
on small scales ($\sim 1-10h^{-1}Mpc$) (for early works, see Totsuji
\& Kihara 1969; Davis \& Geller 1976; Davis \& Peebles 1983). In
recent investigations, the indexes were found to be $ \gamma \sim
1.8$ and $r_0\sim 6.1h^{-1}Mpc$ for the SDSS data and $\gamma \sim
1.7$ and $r_0\sim 5.1h^{-1}Mpc$ for the 2dFGRS data (Zehavi et al.
2002; Hawkins et al. 2003). As pointed out by Yadav et al. (2005),
the power law relation of $\xi (r)$ does not hold on large scales
and it breaks down at $r>16h^{-1}Mpc$ for SDSS and at
$r>20h^{-1}Mpc$ for 2dFGRS, and thus it would not violate the
homogeneous nature of the Universe.

An advantage of the two-point correlation function is its
application to galaxy groups to reveal the matter distribution in
the Universe, since the occupation of haloes by galaxies is believed
to depend on halo mass (see Benson et al. 2000; Berlind et al.
2003). As each dark matter halo is expected to give birth to a
single group of galaxies, applying the two-point correlation
function to different groups of galaxies one can reveal the
clustering nature of these groups and this in turn can tell how dark
matter is distributed (see, e.g., Jing \& Zhang 1988; Merchan et al.
2000; Zandivarez et al. 2003; Padilla et al. 2004).

In measuring the scale of homogeneity of the Universe, the power
spectrum is applicable (see Einasto \& Gramann 1993). Martinez et
al. (1998) introduced the $K$ function $K(r)$, related to the
integral of the two-point correlation function, as a tool. One of
their main claims is that the estimators for $K(r)$ are more
reliable than the most currently used estimators for $\xi (r)$ and
that makes the use of $K(r)$ recommendable (especially in
three-dimensional processes and at large scales) despite its
somewhat less informative character (see Martinez et al. 1998). In
the following one will find that the concept of the $K$ function is
also useful in describing the crowded nature of the neighborhood of
individual sources and hence able to reveal the density property of
a group of sources.

Based on the previous methods, we develop in this paper a
statistical approach to sort out clustering sources. Our main
concern includes: crowded degrees of sources in their neighborhoods
are quantified so that one can tell how different crowded sources
form a cluster; redshift dependent densities of samples are taken
into account to sort out clusters so that the relevant statistical
significance could be determined; via simulation, probabilities for
forming clusters by chance are calculated. The method is applied to
the 2dFGRS data set. As the results, some fine maps of large-scale
structures derived from the sample are available and probabilities
of a list of clusters are obtained. The paper is organized as
follows. In Section 2, we propose the method, where a statistic
called the individual neighborhood function $\kappa _i(r)$ is
introduced to describe the crowded nature of individual sources. In
Section 3, we describe the selection of the sample and study the
redshift distribution it obeys. Density distributions of the sample,
in terms of $\kappa _i(r)$, are presented in Section 4. Clustering
probabilities obtained by simulation are studied in Section 5.
Coherent clusters are identified and shown in Section 6. In Section
7, we show the structures of some large prominent coherent clusters.
In Section 8, the largest structures sorted out from the whole 2dF
data set are shown and discussed. In the last section (Section 9), a
summary of the paper and a brief discussion are presented.


\section{Methods and relevant statistics }\label{sec:blt}

\subsection{The clustering scale }

There are two well-known methods for identifying galaxy groups or
clusters. The most popular one is used to create galaxy group
catalogues, where a projected separation together with a velocity
difference are employed to search for companion sources (Huchra \&
Geller 1982; Eke et al. 2004; Diaz et al. 2005; Merchan \&
Zandivarez 2005). A product of this method is the property
associated with the virial theorem (e.g., the velocity dispersion of
a set of sources), and it is suitable to identify virialized group
of galaxies. The other is used to pick out large-scale structures,
where a single scale, which is applied to all sources of the sample,
is employed to identify a neighborhood source (Einasto et al. 1984;
Deng et al. 2006). The main concern of this method is the
statistical property (e.g., the length or density) of the identified
clusters. In both approaches, the so-called ``friends-of-friends''
(FoF) algorithm, the claim ``any friend of my friend is my friend'',
is applied. The latter approach is somewhat suitable for our
analysis. However, besides picking out clustering sources, we need
to know in what confidence level a group of sources could be
regarded as a cluster or with what probability the cluster could be
identified by chance. Here, based on the second method, we try to
establish a statistical approach to identify large-scale structures,
with which the confidence level of selecting a cluster is able to be
informed.

In order to develop such a method, we need to address several
questions: (1) How do we pick out a cluster? (2) In what confidence
level could a cluster with a certain number of galaxies and crowded
degree be identified by chance? To find answers to these questions,
at least the following requirement should be satisfied: the
probability for identifying a group of galaxies as a cluster by
chance should be well determined. In doing so, it is essential that
all selection effects should be removed or avoided. It is plain
that, if the scale used to find coherent clusters is larger than the
area concerned, then all sources within that area must be included
in a single coherent cluster; if the scale is as small as possible,
then all sources within that area must be separated from each other
and no coherent clusters could be identified. It seems that, in
identifying a group of clustering sources, one needs to take several
steps: determine a criterion clustering scale, $r_{ccs}$; sort out
clustering sources with $r_{ccs}$ by applying the FoF algorithm;
calculate the probability for picking out a cluster by chance. In
calculating the probability, the Monte-Carlo simulation will be
applicable.

In the case of the two-point correlation function, the correlation
length $ r_0$ is regarded as a characteristic clustering scale.
According to the well-known power law relation $\xi
(r)=(r/r_0)^{-\gamma }$, the probability of finding another source,
relative to any source concerned, at $r=r_0$ is twice the
probability produced by random data sets. If the sample is large
enough, the twice probability suggests that the density measured at
that scale is two times of the mean density. As mentioned above, in
the two recent surveys, $r_0$ was found to range from $\sim
5.1h^{-1}Mpc$ to $\sim 6.1h^{-1}Mpc$. A criterion associated with
the sorting out scale comparable to these scales will be adopted in
this paper.

Let $\rho _0(z)$ be the mean density of a flux-limited but
homogenous data set without the effect of clustering, expected at
redshift $z$. Similar to the commonly used selection function (see,
e.g., Martinez \& Saar 2002b; Coil et al. 2004; Padilla et al.
2004), $\rho _0(z)$ as a function of redshift is determined by the
real distribution of the observed galaxies at $z$, which depends on
the evolutionary effect, and is affected by the flux-limited effect
as well as other selection effects such as masks in given fields,
fiber collisions in the spectrographs, etc (Martinez \& Saar 2002b).
As pointed out by Martinez \& Saar (2002b), many selection effects
are directional, with some being due to the construction of the
sample. The absorption by the dust of the Milky Way also gives rise
to a selection effect which is also directional. The best way to
take this effect into account is to consider the well defined maps
of the distribution of galactic dust (Schlegel, Finkbeiner and Davis
1998). When directional selection effects are considered, we should
deal with $\rho _0(z,\theta,\phi)$ instead of $\rho _0(z)$, where
$\theta$ and $\phi$ are coordinates of direction. For the sake of
simplicity, we consider in this paper only the case of $\rho _0(z)$
which corresponds to the commonly used radial selection function.
The difference between $\rho _0(z)$ and the radial selection
function is that, when measuring the former, we divide the count of
sources within a redshift interval by the real volume confined by
that interval.

The criterion clustering scale adopted in this paper is taken as
$r_{ccs}=[2\rho _0(z)]^{-1/3}$. The reason for adopting this
criterion clustering scale is that the scale so adopted is generally
smaller than, but comparable to, the mentioned typical scales $r_0$
which range from $\sim 5.1h^{-1}Mpc$ to $\sim 6.1h^{-1}Mpc$ in the
two recent surveys. The mean of $r_{ccs}$ in the sample adopted
below (sample 2) is $4.2h^{-1}Mpc$. The number of galaxies in the
sample with their $r_{ccs}$ being less than $5h^{-1}Mpc$ is 172265,
76\% of the total. That means that sources of clusters will
generally be sorted out with the scales less than $r_0$ (note also
that, in Deng et al. 2006, $r\sim 5h^{-1}Mpc$ was adopted to sort
out clusters, and in Hoyle et al. 2002, $ r=5h^{-1}Mpc$ was used to
plot the median density contour). We find that when
$r_{ccs}=5h^{-1}Mpc$ is adopted to sort out clusters one will get a
much larger number of galaxies for the largest coherent cluster
sorted out with the method proposed below. Adopting $r_{ccs}=[2\rho
_0(z)]^{-1/3}$ will provides us a conservative result. (Note that
when adopting $r_{ccs}=[4\pi\rho _0(z)/3]^{-1/3}$\ one will get a
more conservative result since $r_{ccs}$ becomes smaller, but this
scale will be much less comparable to $r_{0}$.)

It is known that in the case of the two-point correlation function,
$r_0$ depends on the clustering property of samples. If adopting
$r_0$ to identify clusters of a sample, we are going to measure the
structure of the sample with a ruler concluded from the structure
itself. If the structure changes then the ruler changes. Unlike
$r_0$, the criterion clustering scale $ r_{ccs}$ proposed here
depends only on the property of the background sample from which the
concerned probability will be derived (see what presented below).
This quantity is entirely independent of the properties of the
adopted sample and then is a rather objective ruler.

Due to the flux-limited effect, some distant galaxies will be
missed. This makes the average distance between distant sources that
are observed become larger. If we take a fixed $r_{ccs}$ to identify
clusters by applying the FoF algorithm, we might probably include
all nearby sources in a single large-scale structure when $r_{ccs}$
is large enough, while it might be possible that for those far away
galaxies only very few of them are included to form a cluster since
$r_{ccs}$ is not as large as to connect most of them. This is
unfair. For some nearby sources, they might in fact be relatively
far away from the local structure but they are included as members
of the structure since their distances to the structure are shorter
than the adopted $r_{ccs}$; for distant sources, since the average
distances are large they have less chance to be identified as
members of clusters according to this fixed $r_{ccs}$. When we adopt
$r_{ccs}=[2\rho _0(z)]^{-1/3}$ we deal with a varying $ r_{ccs}$
since $\rho _0(z)$ is a function of redshift. In this case, we will
get a large value of $ r_{ccs}$ for large redshifts and obtain a
smaller one for small redshifts. As part of out method, a varying $
r_{ccs}$ will be adopted to identify clusters (see what proposed
below).


\subsection{Radial selection function }

The result of simulation would be acceptable when the background
sample so created has no bias or has only very insignificant bias.
It is essential that all selection effects and the evolutionary
effect have been removed or avoided.
There is a straightforward technique to avoid these effects. Assume
that, within a spherical space centered at the observer, there are
$N_0$ galaxies in total, which form various structures due to
dynamics and other known or unknown factors. When no clustering
mechanisms are at work, these sources are expected to be
homogeneously distributed within the area, and they constitute a
global homogenous background sample. In fact, what one can observe
at present are galaxies of different cosmological ages. Due to the
evolutionary effect, number densities might evolve with redshift,
and then the total number, $N$, of galaxies whose photons could
reach us at present would be different from $N_0$ (e.g., the
formation of galaxies and the merging of galaxies might occur
somewhere at some time). In this situation, the distribution of the
observable background sample would only be homogeneous locally. Let
us consider a survey over the whole sky. Due to the flux-limited
effect and other selection effects, only $N^{\prime }$ ($N^{\prime
}<N$) galaxies could be observed in both real and background
samples, which we call the whole observed real sample and the whole
observed background sample respectively. (Note that these two
samples have the same number of galaxies, but the galaxies are
assumed to be distributed randomly in the latter sample.) It is
obvious that the mean number densities of the whole observed real
sample and the whole observed background sample would be the same in
different redshifts, from which we can deduced a mean number density
law with respective to redshift. Now we perform a survey over a
limited area of the sky. According to the isotropic nature of the
Universe, it is expected that the mean number densities of this
smaller observed real sample and the corresponding observed
background sample are the same as well in different redshifts.
Keeping this in mind, a background sample created by the simulation
following this mean number density law would have no bias.
Statistical significance derived from this sample would be safely
acceptable. Note that selection effects would differ from survey to
survey due to different techniques and instruments. The proposed
simulation method is simple due to the fact that it does not refer
to the details of selection effects (in fact, all selection effects
have been taken into account).

Besides the selection effects, there exists the boundary effect for
most samples. It is desired that problems arising from this effect
could be overcome. The most effective way of solving this problem is
to consider the real volume involved. In the case of estimating $\xi
(r)$, the count of sources of the adopted sample lying inside the
shell $ [r-\Delta r/2,r+\Delta r/2]$ relative to a point is divided
by the volume of the intersection of that shell with the whole
sample volume (Rivolo 1986; Martinez et al. 1998, 2001). This
technique will be adopted in our analysis.

Based on these arguments, we propose: a) for a survey concerned,
search for the mean number density law $\rho _0(z)$ with the whole
data set; b) draw from the survey a subsample for which the borders
are regularly cut so that within these borders a sub background
sample could be easier created according to $\rho _0(z)$; c) for a
given redshift, determine the sorting out coherent clustering scale
by $r_{ccs}(z)=[2\rho _0(z)]^{-1/3}$; d) apply the FoF algorithm
with $r_{ccs}(z)$ to identify clustering sources; e) calculate the
probability of creating a cluster with a given number of galaxies
and density by chance. In doing so, the real volume confined by $
r_{ccs}(z)$ and the whole area of the subsample will be considered
when calculating the relevant density (see what presented below) so
that the boundary effect would be avoided (see Rivolo 1986; Martinez
et al. 1998, 2001).

What effects does varying $r_{ccs}(z)$ have on the analysis when the
above method is applied? As discussed above, the missing of objects
due to the flux-limited effect will make $\rho _0(z)$ becoming
smaller and thus make $r_{ccs}(z)$ becoming larger. Two
possibilities will occur in this situation: some intrinsic
clustering sources are missed and some intrinsic unclustered objects
are included in a cluster identified with the method. This will make
the sorting out clusters less convincing at large distances where
galaxies are rare. The probability for mis-identifying sources is
currently unavailable since the real distribution of galaxies is
unaware.
%

\subsection{Definition and estimation of the neighborhood function }

The concept of clusters indicates the phenomenon of a number of
sources gathering within a relatively small volume. It is directly
associated with high density regions. The two-point correlation
function is not a direct measurement of number density of the
neighborhood of individual sources and cannot describe the crowded
nature (the degree of density) of their neighborhoods. The same
number found within the same volume of shells relative to two
sources does not guarantee that the same amount of neighborhood
galaxies would be detected within the area enclosed by the shells
(this could be observed in the galaxy distribution plot of many
surveys). To measure the crowded nature of individual sources as
well as the density of a cluster, we need a proper quantity which
could be applicable to a discrete source sample.

There is a quantity suitable to describe the crowded nature of the
neighborhood of discrete sources. It is the conditional density in
spheres $\Gamma^* (r)$ which is an ensemble of realizations of a
given point process (see Coleman et al. 1988; Joyce et al. 1999,
2005; Vasilyev et al. 2006). Motivated by this quantity and the $K$
function mentioned above in the introduction section, we introduce
in the following a new statistic to describe the crowded nature of
the neighborhood of individual sources, where the mean density as a
function of redshift and the boundary effect are taken into account.

Consider a real sample with number of galaxies $N$ and a background
sample of the same number of galaxies, confined within the same
volume. The latter sample is created by simulation under the
assumption that galaxies are randomly distributed according to $\rho
_0(z)$ over the space concerned. Assign $\eta (\overrightarrow{r})$
as a density function defined by
\begin{equation}
\eta (\overrightarrow{r})\equiv \frac{\rho (\overrightarrow{r})}{\rho _0(z)%
}-1
\end{equation}
in the co-moving coordinates of the Universe, where $\rho
(\overrightarrow{r})$ is the number density of the sample concerned
at position $ \overrightarrow{r}$, and $\rho _0(z)$ is the mean
density of the background sample expected at the redshift
corresponding to $ \overrightarrow{r}$. Concerning the crowded
nature of the neighborhood (the neighborhood density) of a source,
let us introduce an individual neighborhood function $\kappa _i(r)$
for object $i$ to describe the number density of the sample relative
to the object within scale $r$, which is defined by
\begin{equation}
\kappa _i(r)\equiv \frac{\int_{V_i(r)}\rho
(\overrightarrow{r})dV}{\rho _0(z_i)V_i(r)}-1,
\end{equation}
where $z_i$ is the redshift of object $i$, and $V_i(r)$ is the
volume of the intersection of the sphere with radius $r$, centered
at object $i$, with the whole volume of the adopted sample (here,
$r$ does not represent $ |\overrightarrow{r}|$). The mean of $\kappa
_i(r)$ of a sample (called the neighborhood function of the sample),
\begin{equation}
\kappa (r)\equiv \frac 1N\sum_i^N\kappa _i(r),
\end{equation}
is one of its statistical properties (a spatial distribution
property), which reflects the mean relative number density of the
sample measured
within the mentioned scale (note that, in the case of adopting $r=r_{ccs}$, $%
r $ will differ from source to source). For a subset of the sample
(namely, a group of sources of the sample), we also use the mean of
$\kappa _i(r)$ to describe its spatial distribution property (its
density nature), and this quantity is also denoted by $\kappa (r)$.
The only difference is that, in equation (3), $N$ will be replaced
by $N^{\prime }$, where $N^{\prime }$ is the number of galaxies of
the subset. It is obvious that the larger the $\kappa (r)$, the
more crowded the sample or the group. In terms of $\eta (\overrightarrow{r})$%
, the two quantities could be expressed by
\begin{equation}
\kappa _i(r)=\frac{\int_{V_i(r)}\eta
(\overrightarrow{r}_i)dV}{V_i(r)}
\end{equation}
and
\begin{equation}
\kappa (r)=\frac 1N\sum_i^N\frac{\int_{V_i(r)}\eta
(\overrightarrow{r}_i)dV}{V_i(r)}.
\end{equation}
(Note that, when different weights for the particles are adopted and
taken
into account, one might deal with other definitions of $\kappa _i(r)$ and $%
\kappa (r)$. In that situation, one might wish to modify definition
(1) so that the forms of other equations would
maintain.)\footnote{For example, when one assumes a real density
$\rho _0(z)$ but deals with a sample for which some galaxies are
missed, one might wish to introduce a weight to account for the
missing sources. In this way, $\rho (\overrightarrow{r})$ would be
replaced by $\rho (\overrightarrow{r})/w(z)$, where $w(z)$ is the
weight which is asumed to be known, and then equation (1) is
modified. Absorbing $w(z)$ into $\rho _0(z)$, $\rho' _0(z)=\rho
_0(z)/w(z)$, and replacing $\rho _0(z)$ with $\rho' _0(z)$, one will
get the same form of equations
for $\kappa _i(r)$ and $%
\kappa (r)$.}

In the case of the two-point correlation function, $\xi (r)>0$
suggests that, for any given source, the probability of finding
another source in distance $r$ is larger than the probability
produced by the random data set and the former is $1+\xi (r)$ times
of the latter. In the case of the individual neighborhood function,
the probability of finding a source within the scale of $r$ relative
to position $\overrightarrow{r_i}$ is proportional to the integral
of the number density $\rho (\overrightarrow{r})$ over the volume
confined by $|\overrightarrow{r}-\overrightarrow{r_i}|<r$ for the
adopted sample, and it is $1+\kappa _i(r)$ times of that of the
background sample which is created randomly according to $\rho (z)$,
where $\kappa _i(r)>0$ suggests the larger probability of the sample
than that of the background sample. In the case of the neighborhood
function, the probability of finding a source within the scale of
$r$ is $1+\kappa (r)$ times of that of the background sample.

In practice, $\kappa _i(r)$ and $\kappa (r)$ can be computed by
\begin{equation}
\kappa _i(r)=\frac 1{\rho _0(z_i)\Delta V_i(r)}\sum_j^{|\overrightarrow{r}%
_j-\overrightarrow{r}_i|<r}\int n_jdV-1
\end{equation}
and
\begin{equation}
\kappa (r)=\frac 1N\sum_i^N[\frac 1{\rho _0(z_i)\Delta V_i(r)}\sum_j^{|%
\overrightarrow{r}_j-\overrightarrow{r}_i|<r}\int n_jdV]-1,
\end{equation}
where $n_j$ is the $\delta $-function giving the position of
particle $j$; $\Delta V_i(r)$ is the portion of
the whole sample volume satisfying $|\overrightarrow{r}-\overrightarrow{r}%
_i|<r$ for any position $\overrightarrow{r}$; the integral takes
over $\Delta V_i(r)$; the sum over $j$ includes only
particles in the position within $|\overrightarrow{r}_j-\overrightarrow{r}%
_i|<r$ in the sample, which yields the number of sources found within $%
\Delta V_i(r)$. For a volume limited sample, $\Delta V_i(r)$ would
differ from source to source due to different positions of the
objects relative to the borders of the sample volume, even $r$ is
the same.

One can check that, when ignoring the variation of $\rho _0(z)$ with
respect to redshift, then quantity $1+\kappa (r)$ would be similar
to the Ripley $K$ function which is an integral of $1+\xi (r) $ over
the spherical ball with radius $r$ (Ripley 1981; Martinez et al.
1998). There is not a direct relation between $\kappa (r)$ estimator
and commonly used $\xi (r) $ estimators since we calculate $\kappa
(r)$ in a slightly different manner [e.g., no random samples are
required to compute $\kappa (r)$]. However, when the difference
between $1+\kappa (r)$ and the Ripley $K$ function is ignored, then
the differential of $1+\kappa (r)$ would come to $1+\xi (r) $. The
main difference between the $K$ function, or the two-point
correlation function, and the neighborhood function is that, with
$\kappa _i(r)$ we can tell how crowded is the neighborhood of a
source and can even show how different crowded sources are
distributed in the space and how this distribution is related to the
structures observed (see what presented below).

Quantity $\kappa (r)$ could also be closely related to the
conditional density in spheres $\Gamma^* (r)$ (Vasilyev et al.
2006). In applying equation (7), when one takes $\rho _0(z_i)=1$ and
makes the sum over the whole sample, $1+\kappa (r)$ would become
$\Gamma^* (r)$, as long as the volume of the sample concerned is
large enough so that for any point in the sample the sphere of
radius $r$ centered at the point is well inside the volume.

In the following analysis, cosmological distances as well as
co-moving volumes will be expressed in terms of the co-moving
coordinates and in calculating the latter the standard cosmological
model, $(\Omega _M,\Omega _\Lambda )=(0.3,0.7)$, will be adopted
through out this paper, leaving the Hubble constant parameter $h$
($H_0=100hkms^{-1}Mpc^{-1}$) serving as a unit. The analysis will be
performed in redshift space. That is, when calculating the co-moving
coordinates of galaxies of samples drawn from redshift surveys, the
peculiar velocity of individual sources will be ignored and the
measured redshift will be taken as the real distant indicator of the
object.

\section{ Sample selections and redshift distributions }\label{s.lpt}

The data studied are taken from ``The 2dF galaxy redshift survey
(2dFGRS): final data release'' (Colless et al. 2003). Two main
regions of the survey data located in distinct areas of the sky are
visible. To calculate $\kappa
_i(r)$ as well as $\kappa (r)$, we must know the volume $%
\Delta V_i$ for each source and for any given $r$. This would be
realizable if the area concerned is well defined where its borders
are sharp. The computation would be easier performed if the edges of
a region are regularly cut. The following criterions are adopted to
select a sample from the survey data set: $150.0^{\circ
}<R.A.(J2000)<220.0^{\circ }$, $-4.8^{\circ }<Dec.(J2000)<1.0^{\circ
}$, $0.0<z<0.3$, $quality\geq 3$, where $quality$ is the redshift
quality parameter for best spectrum ($quality=1-5$; reliable
redshifts have $quality\geq 3$). We thus get a sample of 59497
sources (called sample 1). Table 1 presents the description of this
sample as well other samples mentioned below, where column 1 is the
name of samples, columns 2, 3 and 4 denote the number, area and
region of the samples respectively, column 5 shows the $\rho _0(z)$
that is used to calculate $\kappa$ for the samples and to create the
corresponding background samples by simulation (see what mentioned
below), and column 6 denotes the type of data (observation or
simulation, with the former being that of a real sample and the
latter being that of a background sample). The sky region of sample
1 is illustrated in figure 1, where one of the two distinct areas,
which encloses this sample, is also presented.

We notice that there are shortcomings of the 2dF sample. As shown in
Cannon et al. (2006), there are fibre collisions within 2dF, where
many objects around the edge of a 2 degree field are missed (see
figure 3 in Cannon et al. 2006). This would give rise to a
directional incompleteness of a survey. However, the provided data
are from such a set of 2 degree fields where most regions of the sky
inside the survey boundary are covered by several overlapping fields
and the potential incompleteness has been significantly reduced (see
figure 5 in Colless et al. 2003). In addition, we cut away the edge
of the concerned region and this reduces the directional effects of
fibre collisions as well. We thus presume in this paper that the
effect of this is minor (i.e. ignorable).

Another shortcoming is the existence of ``missed galaxies'' (see
Pimbblet et al. 2001; Cross et al. 2004). As shown in Cross et al.
(2004), the incompleteness of the sample could reach about 14 per
cent, which varies slightly with magnitude (see figure 11 of the
paper). This shortcoming is a selection effect. As a function of
redshift, it is absorbed into $\rho _0(z)$ (see what adopted below).
The main influence of this effect on our results is that when we
identify a large-scale structure with the current 2dF data we might
obtain a larger one when all missed galaxies are discovered.

The third shortcoming is due to the fact that the $b_{J}=19.45$
magnitude limit is not constant across the entire area of the survey
(see Colless et al. 2001; Norberg et al. 2002). This might affect
the results of the analysis. We will discuss this effect in the last
section of this paper.

\begin{figure}[tbp]
\begin{center}
\includegraphics[bb=0 0 560 460]{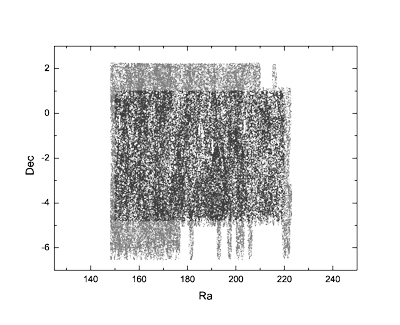}
\end{center}
\caption{Sky region of sample 1 (dark grey color), in coordinates of
$R.A.(J2000)$ and $Dec.(J2000)$. The parent area where sample 1 is
taken from is also presented (both dark and light grey colors).}
\label{Figure 1}
\end{figure}

Since the volume of the area concerned is well defined, we are now
able to estimate the mean density of the background sample, $\rho
_0(z)$, from the observational data. As pointed out by Padilla et
al. (2004), there are several approaches to do so. The most
straightforward way is to use the luminosity function of galaxies to
obtain an estimate of $\rho _0(z)$. This is not adopted in this
paper since any bias from the real luminosity function would cause
systematic errors (see Padilla et al. 2004 and the references
therein). In the following, let us consider three other approaches.
The first way is to estimate $\rho _0(z)$ directly from the
observational data. In this approach we assume that the redshift
distribution of the whole 2dF survey data set could serve as a
parent population of the background sample. Of course, to match the
data of sample 1, the data set serving as the parent population,
which covers the whole area of the 2dF survey, should meet these
criterions: $0.0<z<0.3$, $quality\geq 3$. A set of 226302 sources
(called sample 2) is obtained, which is almost four times of the
number of galaxies of sample 1. As illustrated below, there are two
massive superclusters in 2dF and these superclusters are comparable
to the survey volume (see also Erdogdu et al. 2004; Lahva and Suto
2004; Porter \& Raychaudhury 2005; Einasto et al. 2006a). This
suggests that 2dF might not be a fair sample of the Universe. The
reason for assuming the whole 2dF survey data set being a fair
sample of the Universe is that it is indeed very large (based on it,
many statistical analyses have been made) and it is the largest
available sample of this kind (detected by the same means of the 2dF
survey). Under this assumption, we simply measure $\rho _0(z)$ from
sample 2 and then apply it to our simulation analysis. The advantage
of this approach is that any possible evolutionary and selection
effects, known or unknown, will be accounted for. There are two
disadvantages of the method. One is due to the measurement
uncertainty which will cause an unreal distribution within the
uncertainty scale. This will be eased by assigning a random value
smaller than the uncertainty to the background data when randomly
drawing them from sample 2 (the following application shows that
this technique is indeed at work; see figure 3). The other is due to
the fact that the 2dF survey does not covers the whole sky. If there
exist strong clusters which are comparable to the regions concerned
(yes, as pointed out above, there are some), the redshift
distribution of the presumed parent population will be affected:
bumps will be seen in the very dense place and troughs will exist in
the very sparse space. In this situation, some redshift distribution
feature due to clustering (Padilla et al. 2004) will be mistaken as
the background property and then clusters corresponding to this
feature might be missed (at least in the redshift space this
clustering property will be ignored). Thus, the number of galaxies
of some very large clusters would become smaller. To get a safer
number of galaxies, one might prefer this approach.

The second way of estimating $\rho _0(z)$ is to fit parametric forms
to the observational data (see Padilla et al. 2004), which uses
sample 2 as well. In this approach, we do not directly apply $\rho
_0(z)$ that is measured from sample 2 to create background samples.
Instead, we fit it with an empirical curve. As long as the
probability obtained from the fit is small enough, the fitting curve
will serve as $\rho _0(z)$ and then will be applied to our
simulation analysis. The mean density curve so obtained is nothing
but a deep smoothing of that directly measured from sample 2. We
have a risk in employing this approach. In the process of smoothing,
while bumps and troughs caused by clustering will be smoothed,
features arising from selection effects will be smoothed as well. In
the following analysis, results arising from the first and the
second approaches will be presented and compared.

For the sake of comparison, the third way of estimating $\rho _0(z)$
is also adopted. In this approach, the galaxies observed are assumed
to be homogenously distributed over the whole area of sample 1. Thus
we simply take $\rho _0(z)=const$ to create background samples.
Although the redshift distribution of this kind of background sample
is not real, but the analysis might reveal some aspects of
clustering which are unfamiliar before (see what presented below).

Let us derive $\rho _0(z)$ from sample 2. The count observed within
any sky region of the 2dF survey is assumed to be proportional to
the corresponding sky area according to the isotropic property of
the Universe. Let $z_i$ be the redshift of source $i$ in sample 2.
We calculate the count within the redshift interval $0.99z_i$ ---
$1.01z_i$ from sample 2 as well as calculate the volume $V(\Delta
z_i)$ confined by this redhift interval in the area of sample 1.
According to the isotropic assumption, the volume confined by this
redshift interval in the whole area of sample 2 should be $
226302/59497$ times of $V(\Delta z_i)$. Thus the count divided by $
(226302/59497)V(\Delta z_i)$ will be taken as the estimated value of
$\rho _0(z)$ for this source, which is denoted by $\rho _0(z_i)$. In
doing so, when the redshift interval is less than the measurement
uncertainty of the survey, the latter will be taken as the width of
the redshift interval.

\begin{figure}[tbp]
\begin{center}
\includegraphics[bb=0 0 660 560]{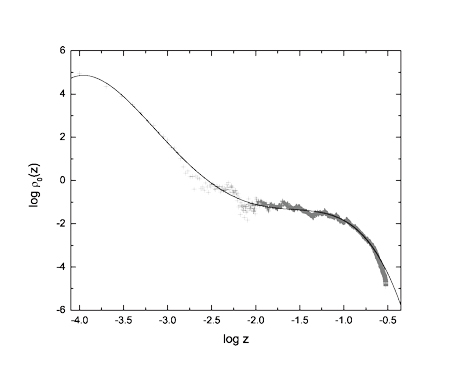}
\end{center}
\caption{Relation between the mean number density of sample 2
(pluses) and redshift, where the solid line is the fitting curve of
the data.} \label{Figure 2}
\end{figure}

The estimated values of $\rho _0(z_i)$ of sample 2 are shown in
figure 2. The data could be well fitted by the following polynomial
function: $\log \rho _0(z)=-(11.2931\pm 0.0049)-(20.585\pm
0.013)\log z-(15.587\pm 0.011)(\log z)^2-(4.9779\pm 0.0039)(\log
z)^3-(0.52885\pm 0.00045)(\log z)^4$ ($P<0.0001$ ), where $\rho
_0(z)$ is in units of $(Mpc/h)^{-3}$.

We create a background sample (sample 3) randomly from sample 2 with
the first approach. The second background sample (sample 4) is
produced with the second approach by simulation according to the
empirical mean number density function, the polynomial function. The
third background sample (sample 5) is yielded by simulation under
the assumption that the galaxies are homogenously distributed, where
the mean number density is obtained by dividing 59497 with the whole
volume of sample 1, which gives $\rho _0=0.00247(Mpc/h)^{-3}$. The
numbers of galaxies of all these background samples are the same as
that of sample 1, $N=59497$.

\begin{figure}[tbp]
\begin{center}
\includegraphics[bb=0 0 660 600]{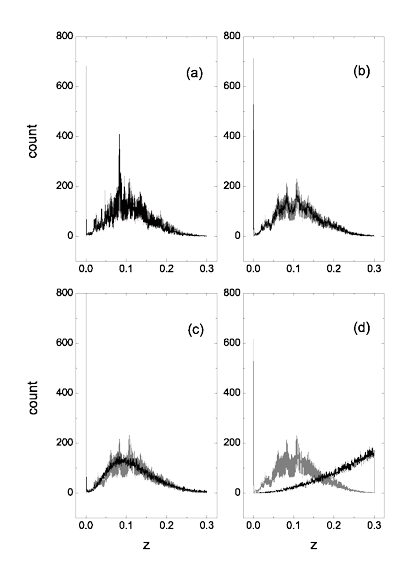}
\end{center}
\caption{Redshift distributions of samples 1 (black lines in panel
a), 2 (grey lines in all panels), 3 (black lines in panel b), 4
(black lines in panel c), and 5 (black lines in panel d). Here, the
number of galaxies of sample 2 is normalized to that of sample 1.}
\label{Figure 3}
\end{figure}

Redshift distributions of samples 1-5 are shown in figure 3. Panel
(a) of the figure shows that there indeed exist some bumps and
troughs in both samples 1 and 2. Compared with other panels, we
suspect that this phenomenon is unlikely to be caused by random.
Instead, it might probably be due to intrinsic distributions such as
large-scale structures (see what discussed above in this section and
see Fig. 22 presented below) or selection effects. In addition, we
find some significant features in sample 1, which are much less
prominent in sample 2. This must be due to the local property of
redshift distribution and might probably be a consequence of the
existence of local clustering (see Figs. 17, 20 and 21 presented
below). Panel (b) plainly demonstrates that samples created with the
first approach well follow the redhsift distribution of sample 2. In
addition, it reveals that the chaos of the observational samples is
largely caused by the measurement uncertainty (since adding an
additional random value smaller than the uncertainty significantly
reduces the fluctuation observed, as panel b shows). From panel (c)
we observe that, when ignoring the fluctuation, which might be
caused by local clustering, the polynomial function could indeed
well represent the real distribution of the observational samples.
Panel (d) illustrates that, as expected, the sources observed are
not homogenously distributed at all (the flux-limited effect and the
evolutionary effect would be the main factors accounting for the
non-homogenous distribution observed in the survey data).

Although spatial distributions of sources of background samples 3
and 4 differ significantly from that of the real homogenous one,
sample 5, a distribution in agreement with the former is still
regarded as a homogenous one in the following analysis. This is
because that background samples 3 and 4 suffer from the flux-limited
effect and the evolutionary effect as well as other known or unknown
selection effects, and it is these effects that lead to the apparent
in-homogenous distribution. When all galaxies of the same
cosmological age are available, the spatial distribution of
background sample sources, which are randomly created, must be
homogenous due to the principle of cosmology. Therefore, in the
analysis below, those in agreement with the spatial distribution of
sources of the adopted background sample, or in agreement with that
of the parent population of background samples, will be regarded as
an intrinsic homogenous one, and those obviously deviating from that
distribution will be considered as an in-homogenous one.

\begin{figure}[tbp]
\begin{center}
\includegraphics[bb=0 0 660 480]{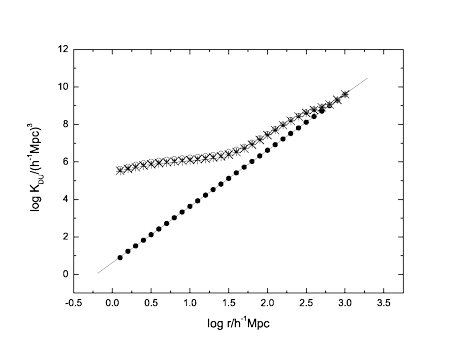}
\end{center}
\caption{Plot of standard $K$ functions estimated from samples 1
(open circles), 3 (pluses), 4 (crosses), and 5 (filled circles). The
solid line is the best linear fit to the $K$ function of sample 5,
which obeys $log K_{DU} = (0.623 \pm 0.003) + (2.996 \pm 0.002) log
r $. The correlation dimension could be easily figured out from the
plot, where $D_2=3$ is obtained for sample 5 and $D_2<3$ could be
derived from the other three samples within the maximum scale of the
samples, $r_{max}=837h^{-1}Mpc$.} \label{Figure 3}
\end{figure}

Shown in figure 2, one could observe a deviation, of the polynomial
provided for $\rho _0(z)$, from the mean number density function
directly measured from sample 2. In addition, a deviation of the
redshift distribution of the polynomial function from that of the
$\rho _0(z)$ of sample 2 is seen in figure 3 (see panels b and c).
This suggests that the polynomial provided for $\rho _0(z)$ and the
$\rho _0(z)$ measured from sample 2 are not the same. However, as
one will see below, this difference does not cause big problems in
the clustering analysis performed in this paper. Since sample 2
suffers from the clustering effect, as mentioned above and revealed
below, it might be possible that samples 3 and 4, which come from
sample 2, would be inhomogeneous not only due to the flux limit but
also due to clustering. Thus, if the deviation observed in figure 2
leads to an observable difference in the homogeneity properties of
the two corresponding background samples needs to be checked. Let us
check this by examining their $K$ functions, from which the
corresponding correlation dimension $D_2$ could be derived (see
Martinez et al. 1998). The $K$ function adopted to investigate this
issue is that provided in equation (8) of Martinez et al. (1998),
the standard estimator introduced by Doguwa and Upton (1989) to
account for the boundary effect. As is generally known, regardless
of the possible slight bias, this estimator, $K_{DU}$, has good
properties. Displayed in figure 4 one could find the $K$ functions
estimated from samples 3 and 4, where, for a comparison, that from
samples 1 and 5 are also presented. As expected, the $K$ function
estimated from sample 5 well follows the homogeneity curve, the
curve for which $D_2=3$. The three other samples are seen not to be
homogeneous at all (for all of them, $D_2<3$ holds within the
maximum scale of the samples). It is interesting that the $K$
functions estimated from samples 1, 3 and 4 follow almost the same
trend (in this case, they will have almost the same $D_2$). While
that from sample 1 is slightly larger than that of samples 3 and 4,
the difference between the latter two cannot be detected. This
suggests that, the $K$ function, and hence the derived $D_2$,
suffers mainly from the redshift distribution which is influenced by
the flux limit, the possible evolutionary effect, and the spatial
distribution of large-scale structure members (see Fig. 3). The role
that the locally clustering (sources in sample 1 are strongly
clustered while those in samples 3 and 4 are not; see the spatial
clustering plots presented below) plays in producing the correlation
dimension $D_2$ seems very insignificant.

\section{ Density distributions }\label{s.nl}

Here, we use estimator (6) to calculate the individual neighborhood
function of sources in the scale of $r_{ccs}$ (i.e., we take
$r=r_{ccs}$) which is determined by $ r_{ccs}=[2\rho _0(z)]^{-1/3}$.
We apply the $\rho _0(z_i)$ directly measured from sample 2 to
calculate $\kappa _i(r_{ccs})$ for each source of samples 1 and 3.
In addition, we determine $\rho _0(z)$ by the polynomial function
obtained above and then calculate $\kappa _i(r_{ccs})$ for each
source of samples 1 and 4 with this $\rho _0(z)$. Also, we adopt
$\rho _0=0.00247(Mpc/h)^{-3}$, the mean number density of sample 1,
to calculate $ \kappa _i(r_{ccs})$ for each source of samples 1 and
5. In this way, we calculate $ \kappa _i(r_{ccs})$ for sample 1 with
three kinds of $\rho _0(z)$ (and hence with three kinds of
$r_{ccs}$) estimated with three approaches.

\begin{figure}[tbp]
\begin{center}
\includegraphics[bb=0 0 660 480]{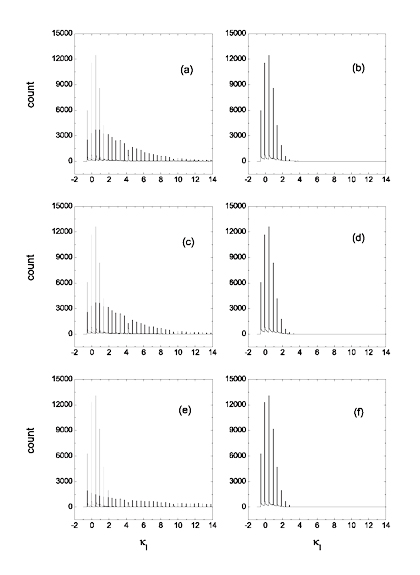}
\end{center}
\caption{Distributions of the individual neighborhood function for
samples 1 (black lines in panels a, c, and e), 3 (black lines in
panel b; light grey lines in panel a), 4 (black lines in panel d;
light grey lines in panel c), and 5 (black lines in panel f; light
grey lines in panel e). The individual neighborhood functions
presented in panels (a) and (b) are calculated with the values of
$\rho _0(z)$ directly estimated from sample 2 (the first
approach); those in panels (c) and (d) are calculated with the values of $%
\rho _0(z)$ determined by the empirical polynomial function (the
second approach); and that in panels (e) and (f) are calculated with
the constant mean density $ \rho _0=0.00247(Mpc/h)^{-3}$ (the third
approach).} \label{Figure 4}
\end{figure}

Distributions of the individual neighborhood function $\kappa
_i(r_{ccs})$ of sample 1, calculated with three kinds of $\rho
_0(z_i)$ are presented in panels (a), (c) and (e) of figure 5
respectively. Meanwhile, distributions of $\kappa _i(r_{ccs})$ of
samples 3, 4 and 5 are displayed in panels (b), (d) and (f) of the
figure, respectively. One might observe from panels (b), (d) and (f)
that the distributions of $\kappa _i(r_{ccs})$ are almost the same
for background samples created with different approaches, although
the adopted $\rho _0(z)$, and hence $r_{ccs}$, differs in different
cases. This is not surprising since $\kappa _i(r_{ccs})$ reflects
only a relative density and background samples are randomly created
by simulation, which would be locally homogeneous. In fact, the
number density of a background sample is created in accordance with
a provided mean density $\rho _0(z)$, and quantity $ 1+\kappa
_i(r_{ccs})$ is that number density divided by the mean density. In
the case of a real sample, the situation is much different. In panel
(e), when calculating $\kappa _i(r_{ccs})$, the number density of
sample 1 is divided by a constant mean density, $\rho
_0=0.00247(Mpc/h)^{-3}$, but the real number density of sample 1
differs significantly for sources with different values of redshift.
Thus, one observes a much different $\kappa _i(r_{ccs})$
distribution of sample 1 in panel (e) from those in panels (a) and
(c). The values of $ \kappa _i(r_{ccs})$ could be found as large as
several ten thousands (the figure covering the whole range of value
of the quantity is omitted). In fact, $\kappa _i$ in panel (e)
reflects absolute number densities rather than locally relative
ones. Due to several selection factors mentioned above, there is an
enormous difference of observed density.
In contrast, distributions of $\kappa _i(r_{ccs})$ presented in
panels (a) and (c) are almost the same. This suggests that the
difference of $\rho _0(z)$ shown in figure 2 does not significantly
influence the distribution of $ \kappa _i(r_{ccs})$ for the adopted
real sample, sample 1. (See also the discussion below in the case of
spatial distributions.)

One might observe that, as histogram plots, distributions in figure
3 look like ``continuums'' since the total number involved is very
large. However, with the same number, those in figure 5 do not look
like ``continuums'' but look like ``histograms''. We have checked
that this difference is not due to the adopted bins. The
distribution plots shown in figure 5 look like ``histograms''
because in some separated ranges of $\kappa _i(r_{ccs})$ the counts
are relatively small, while in the nearby $\kappa _i(r_{ccs})$
ranges the corresponding counts are large (thus, the former counts
look like to be zero). This phenomenon is due to the small scale we
adopt (i.e., $r_{ccs}$). Since the scale is relatively small (see
the definition of $r_{ccs}$ presented above), one could only find
small numbers of sources within the corresponding volume. Thus, the
difference of number would lead to obvious difference of the
individual neighborhood function. For example, the volumes confined
by $r_{ccs}$ would differ slightly for two closely located sources.
If the number of neighborhood galaxies relative to one source is 3
and that relative to the other is 4, then this difference would
cause a 30 percent of change (in this situation it would be hard to
get 3 percent of change for this two sources). At least within a
small region enclosing these two sources, the change of a smaller
percent would not be observed from the difference of numbers other
than 3.

For the background samples, the individual neighborhood function is
mainly distributed within the range of $\kappa _i(r_{ccs})<3$ (see
panels b, d and f), while for sample 1, a large amount of sources
have their $\kappa _i(r_{ccs})$ larger than $3$ (see panels a, c and
e).

Motivated by figure 1 of Governato et al. (1998), let us symbolize a
region within which the initial matter forms a galaxy in later times
with a presumed galaxy (the matter is referred to as the presumed
galaxy matter) and assign the central position of the region as the
position of that presumed galaxy. These presumed galaxies are
expected to be randomly scattered in space. In terms of statistics,
they are considered as un-cluttering sources. In our analysis,
$\kappa _i(r_{ccs})$ distributions of unclustered galaxies are those
displayed in panels (b), (d) and (f) of figure 5. The $\kappa
_i(r_{ccs})$ distribution of sample 1 is obviously different from
them (see panels a, c and e). Revealed in the figure, a large amount
of sources in sample 1 possess very large values of $\kappa
_i(r_{ccs})$ (say, $\kappa _{i}>3$) which the majority of
unclustered galaxies
do not have, suggesting that there is clustering in this sample.

\begin{figure}[tbp]
\begin{center}
\includegraphics[bb=0 0 460 560]{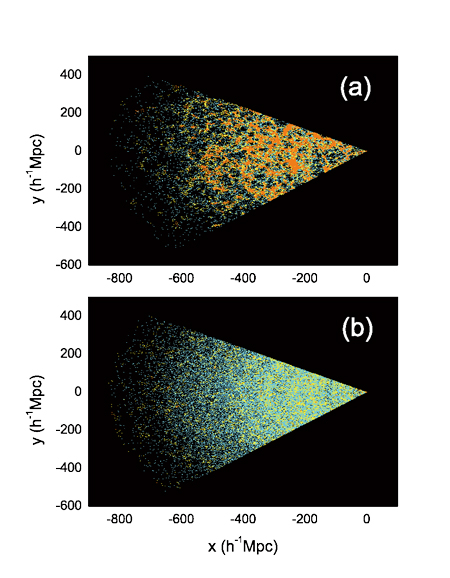}
\end{center}
\caption{Spatial (two dimensional) distributions of different
crowded sources in samples 1 (panel a) and 3 (panel b), measured in
the criterion clustering scale $r_{ccs}$. The values of $\rho _0(z)$
directly estimated from sample 2 are adopted to determine $r_{ccs}$
and to calculate $ \kappa _{i}$. The cyan color stands for
un-crowded sources (sources with relatively low number densities in
their neighborhoods; $-1 \leq \kappa _{i} < 1$) (a lot of cyan color
objects are not observed in this figure due to the overlap of other
color objects), the yellow color represents crowded sources (sources
with relatively large number densities in their neighborhoods; $1
\leq \kappa _{i} < 3$), and the orange color symbolizes very crowded
sources (sources with relatively very large number densities in
their neighborhoods; $3 \leq \kappa _{i}$). The co-moving
coordinates are determined by $x=D sin(\pi/2-Dec) cos(Ra)$, $y=D
sin(\pi/2-Dec) sin(Ra)$, $z=D cos(\pi/2-Dec)$ (here $z$ is a
coordinate, as generally adopted; when it represents a redshift in
the text below, we will present a note), where $D$ is the distance
of the object to the observer, in units of $h^{-1}Mpc$.}
\label{Figure 6}
\end{figure}

\begin{figure}[tbp]
\begin{center}
\includegraphics[bb=0 0 460 560]{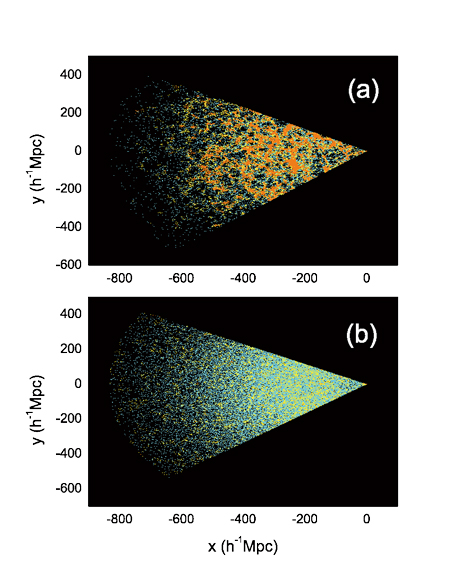}
\end{center}
\caption{Spatial (two dimensional) distributions of different
crowded sources in samples 1 (panel a) and 4 (panel b), measured in
the criterion clustering scale $r_{ccs}$. The values of $\rho _0(z)$
figured out from the empirical polynomial function are adopted to
determine $r_{ccs}$ and to calculate $ \kappa _{i}$. The symbols are
the same as they are in figure 6. For the definition of the
coordinates, see figure 6.} \label{Figure 7}
\end{figure}

\begin{figure}[tbp]
\begin{center}
\includegraphics[bb=0 0 460 560]{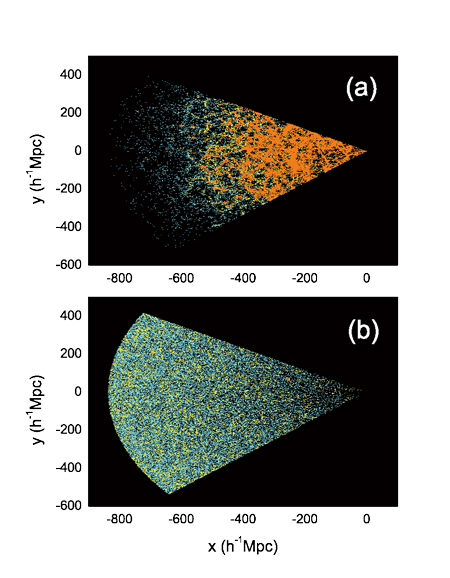}
\end{center}
\caption{Spatial (two dimensional) distributions of different
crowded sources in samples 1 (panel a) and 5 (panel b), measured in
the criterion clustering scale $r_{ccs}$. The mean density $\rho
_0=0.00247(Mpc/h)^{-3}$ evaluated from sample 1, which is a constant
with respect to redshft, is adopted to determine $r_{ccs}$ and to
calculate $ \kappa _{i}$. The symbols are the same as they are in
figure 6. For the definition of the coordinates, see figure 6.}
\label{Figure 8}
\end{figure}

\begin{figure}[tbp]
\begin{center}
\includegraphics[bb=0 0 460 560]{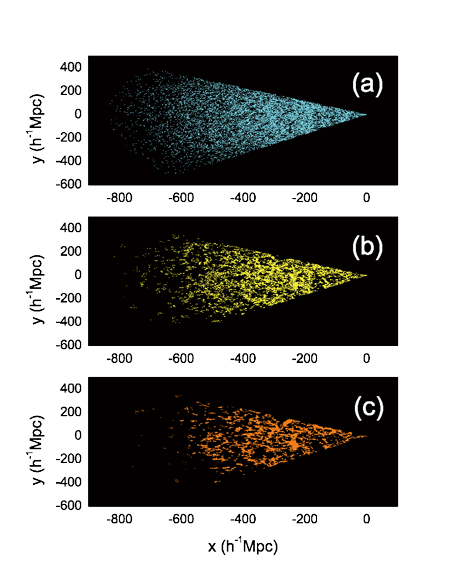}
\end{center}
\caption{Spatial (two dimensional) distributions of un-crowded
(panel a), crowded (panel b) and very crowded (panel c) sources in
sample 1, measured in the criterion clustering scale $r_{ccs}$. This
plot is a copy of panel (a) in figure 6, where different crowded
sources are presented in different panels.} \label{Figure 9}
\end{figure}

\begin{figure}[tbp]
\begin{center}
\includegraphics[bb=0 0 460 560]{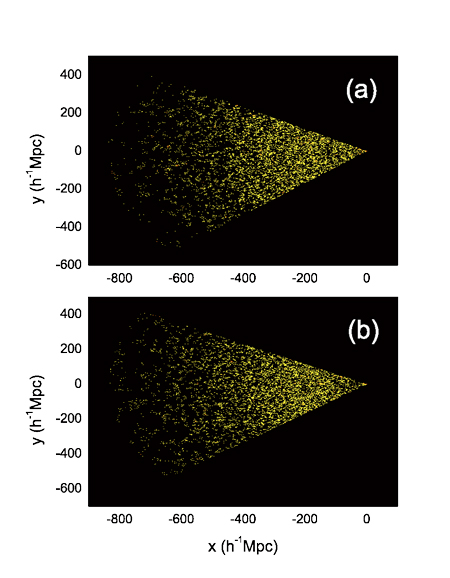}
\end{center}
\caption{Spatial (two dimensional) distributions of crowded and very
crowded sources in samples 3 (panel a) and 4 (panel b), measured in
the criterion clustering scale $r_{ccs}$. Panel (a) in this figure
is a copy of panel (b) in figure 6, while panel (b) in this figure
is a copy of panel (b) in figure 7, where the un-crowded sources are
omitted.} \label{Figure 10}
\end{figure}

According to figure 5 (compare the right- and left-hand-side
panels), we define those galaxies with $-1 \leq \kappa _{i} < 1$, $1
\leq \kappa _{i} < 3$ and $3 \leq \kappa _{i}$ as un-crowded,
crowded and very crowded sources, respectively. In other words,
sources with $-1 \leq \kappa _{i} < 1$, $1 \leq \kappa _{i} < 3$ and
$3 \leq \kappa _{i}$ are identified as those locally located in
regions with relatively small, large and very large number densities
of galaxies, respectively. They have different relative number
density environments. The spatial distributions of different crowded
sources for samples 1, 3, 4, and 5 are displayed in figures 6-8. As
expected, one can find from panel (a) of figures 6 and 7 that very
crowded galaxies seem to form the core of large-scale structures,
where surrounding the core are those less crowded sources (for
identifying large-scale structures, see the analysis below).
Although in the case of adopting the mean number density, the number
of very crowded sources identified from sample 1 (panel a in figure
8) is very large (45753 sources, 77\% of the total number of the
sample) due to the smaller value of the density, $\rho
_0=0.00247(Mpc/h)^{-3}$, the region these sources occupy is smaller
than that in other cases. (In panel a of figures 6 and 7, the
numbers of very crowded galaxies are 27312 and 27476, respectively.)
Also as expected, all the un-crowded, crowded and very crowded
sources in the background samples are seen to be homogenously
(relative to the expected observed density $\rho _0(z)$) distributed
over the whole sample volume (see panel b in the three figures) (a
similar issue will be investigated below; see figure 16). We find
that un-crowded sources in sample 1 show a homogenous distribution
over the whole sample volume, which could be clearly seen in figure
9 (panel a) (see also figures 15 and 16 presented below) (note that
many distant sources are missed due to the flux limited effect). It
suggests that some isolated resided matter regions remain to be
isolated from the very beginning to late times (during this period,
they form galaxies themselves). Crowded and very crowded galaxies in
sample 1 are much less homogenously distributed (see panels b and c
in figure 9). This might be due to the continuous action of gravity
that pulls these galaxies together, as expected in many models.
Governato et al. (1998) have already pointed out that large
concentrations are common in a universe dominated by cold dark
matter and they are the progenitors of the rich galaxy clusters seen
today. It is interesting that filamentary features are observed in
the state of random distribution, formed particularly by crowded
sources (see panel b in figures 6 and 7). This is plainly shown in
figure 10. In addition, seeds of knots seem to be presented early in
the random state as well (see figure 10). According to the view of
Governato et al. (1998), we suspect that it might be these features
that grow to the obvious clustering structure observed in later
times, where many un-crowded presumed galaxies have joined. It seems
that, in this process, very dense regions move towards each other,
and some of them form large-scale structures themselves in later
times. This is just the case expected in the concentration process
of dark matter (see figure 1 of Governato et al. 1998). Revealed by
recent observations, filamentary large-scale structures do exist at
redshift as large as $z\sim 6$ (Ouchi et al. 2005).

\begin{figure}[tbp]
\begin{center}
\includegraphics[bb=0 0 460 560]{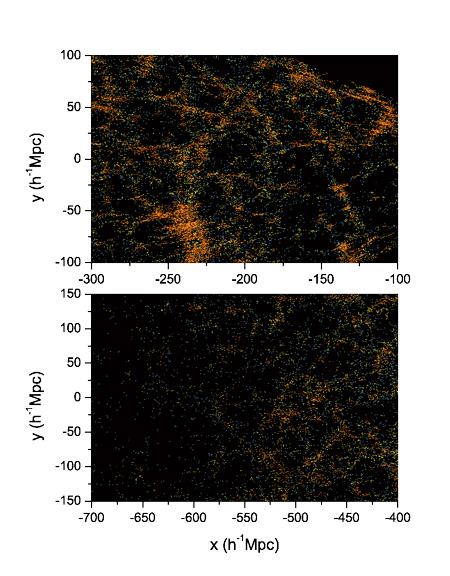}
\end{center}
\caption{Spatial distributions of different crowded sources of
sample 1 in two local regions of the whole volume of the sample (see
figure 6 panel a), where the upper panel presents the plot in a
region closer to us while the lower panel shows the map in a distant
region. The density symbols are the same as they are in figure 6.
For the definition of the coordinates, see figure 6.} \label{Figure
11}
\end{figure}


Observed in the two dimensional map of very crowded sources in
sample 1 are some voids (see figure 9 panel c). These voids are also
detected in the spatial distribution of crowded sources in the
sample, but they are less obvious (see figure 9 panel b). On the
contrary, we find no obvious voids in the spatial distribution of
un-crowded galaxies in the same sample. Meanwhile, some embryonic
voids are seen in the two dimensional plot of crowded sources in
sample 3 (see figure 10 panel a). It suggests that voids are likely
the volumes within which no or very few very crowded sources are
present and they are likely formed in embryo by fluctuation in the
very early epoch of the Universe. It might be the continuous
gravitation in later times that pulls more crowded galaxies closer
and at the same time leaves behind adult voids. Shown in figure 11
are the finer resolution maps showing the details of the spatial
distributions of different crowded sources in sample 1 in two local
regions of the volume concerned (see figure 6 panel a). Indeed, from
the figure we find that, if voids are identified according to the
spatial distribution of very crowded sources, then they are seen to
be filled mainly with un-crowded sources (for a more detail, see
figure 21 presented below).

The spatial density distributions of sample 1 calculated in the
first and second approaches are almost the same (see panel a in
figures 6 and 7). Meanwhile, one might observe that the
distributions of $\kappa _i(r_{ccs})$ for sample 1 calculated with
the two approaches are not distinguishable (see panels a and c in
figure 5). This suggests that the adopted values of $\rho _0(z)$
estimated directly from sample 2 and evaluated by the empirical
polynomial function do not provide a significant difference in the
analysis of density distribution. As discussed above, in examining
homogeneity properties of background samples, where the values of
the correlation dimension $D_2$ are checked, the same conclusion is
obtained. In the following, we discuss only the case of adopting the
$\rho _0(z)$ directly estimated from sample 2 (the first approach),
where radial selection effects (known and unknown) are accounted
for.

\section{Clustering probabilities}\label{s.mf}

As long as the mean density $\rho _0(z)$ is available, we are now
able to apply the FoF algorithm to identify clustering sources with
the scale of $ r_{ccs}(z)$. To calculate the probability of creating
a cluster with a given number of galaxies (in terms of statistics,
the total number of members of a sample is always referred to as the
size of the sample) and with a certain number density by chance, we
perform a number of simulations and for each time of simulation we
sort out coherent clusters and calculate their neighborhood
functions. Let the number of simulations when clusters with numbers
of galaxies larger than $n^{\prime }$ and neighborhood functions
larger than $ \kappa ^{\prime }$ are observed be $\Delta N$, and the
total number of trial be $N$. Then the probability of creating a
cluster with its number of galaxies larger than $n^{\prime }$ and
its neighborhood function larger than $\kappa ^{\prime }$ by chance
is estimated by dividing $\Delta N$ with $N$: $P\{n>n^{\prime
},\kappa >\kappa ^{\prime }\}=\Delta N/N$.

The details of our simulation analysis are summed below (some of
them could be found in the previous sections). a) We randomly select
59497 redshifts from sample 2. These form a simulation sample which
has the same number of sample 1. In doing so, 226302 redshifts which
correspond to 226302 sources contained in sample 2 have the same
chance for being selected. (In this way, the redshift distribution
of the simulation sample would be the same as that of sample 2. See
figure 3 panel b.) b) For each redshift that is selected, we assume
that the digital number that is un-measurable is uniformly
distributed within the range of that digital number and then we
modify the redshift according to this assumption. For instance, when
selecting a redshift of 0.2981 from sample 2, we assume that this
observed redshift arises from a real redshift ranging within
$0.29805<z<0.29815$, and then we modify the original redshift 0.2981
by adding a random value $x$ to it, where $x$ is assumed to be
uniformly distributed within the range of $-0.00005<x<0.00005$. We
then get a new redshift set for the simulation sample. (As shown in
figure 3 panel b, the distribution of redshifts of this kind of
sample is much less scattered than that of sample 2.) c)
Corresponding to each redshift of the simulation sample, we randomly
select a sky position confined in the area of sample 1 and assign
the redshift and the position to a source. In this way, we get 59497
sources for the simulation sample. In doing so, the Universe is
assumed to be isotropic and then the same area in the sky within the
sample region (the area of sample 1; see figure 1) has the same
probability for being selected. d) We calculate $r_{ccs}$ for each
source in the simulation sample by applying the relation $
r_{ccs}=[2\rho _0(z)]^{-1/3}$, where the adopted $\rho _0(z_i)$ is
that directly measured from sample 2. e) For this simulation sample,
we apply the FoF algorithm with the $r_{ccs}$ obtained above to sort
out clusters. For each pair of sources we always have two values of
$r_{ccs}$. The two sources are considered to be within a same
cluster when the distance between them is smaller than the smaller
value of the two $r_{ccs}$. f) For each cluster, we calculate their
$\kappa (r_{ccs})$, using estimator (7), where the sum is taken over
the whole number of sources of the cluster concerned. g) We perform
a number of simulation and then, according to these simulations,
estimate the probability for forming a cluster with a certain number
of galaxies and certain value of $\kappa (r_{ccs})$ by chance.

Besides the reason proposed above, there is one more reason for
choosing $-0.00005<x<0.00005$ as the range of $x$ to be added to the
provided redshift of sample 2. That is, the treatment itself makes
the redshift distribution of the selected sample much less scattered
than that of sample 2 (see figure 3 panel b). In this way, the chaos
caused by the un-measurable digital number of the redshifts is
eased. In doing so, we do not use the real redshift errors since
they are not taken into account when we deal with sample 1. In fact,
as shown by panel (b) of Figs. 6 and 7, spatial distributions of
samples 3 and 4 are not distinguishable, although their redshift
distributions are quite different in finer redshift intervals (see
figure 3). One can check that replacing $-0.00005<x<0.00005$ with
the real redshift errors in the simulation analysis would not
provide an observable difference.

\begin{figure}[tbp]
\begin{center}
\includegraphics[bb=0 0 460 560]{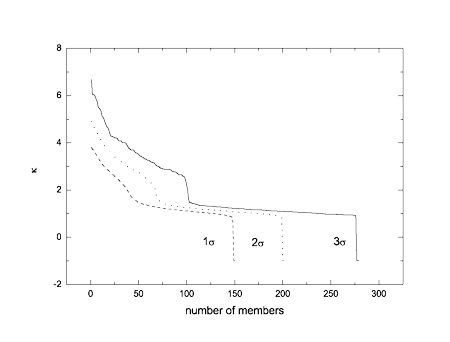}
\end{center}
\caption{Map of probability contours of coherent clusters formed by
chance,
where the dashed, dotted, and solid lines represent the probabilities of $%
1\sigma$, $2\sigma$, and $3\sigma$, respectively.} \label{Figure 12}
\end{figure}

Displayed in figure 12 are the probability contours of coherent
clusters formed by chance (here, $N=5000$). It shows that clusters
with larger numbers of galaxies and larger number densities (larger
$\kappa $) are harder to be formed. The probability of forming
coherent clusters with their numbers of galaxies larger than 350 is
less than that of $3\sigma $, $P(3\sigma )$.

Our analysis shows that, if a cluster with its number of galaxies
$n$ larger than 10 and its $\kappa$ larger than 5.41 could be formed
by chance, the probability must be less than that of $3\sigma$,
$P(3\sigma )$. Under the same requirement of probability, other
typical pairs of $(n,\kappa)$ are $(100,2.24)$, $(200,1.10)$ and
$(300,-0.99)$. It shows that, for a cluster with very large number
of galaxies (say, $n>300$), the requirement of $\kappa$ is very
weak. It could be a negative one. This suggests that clusters with
very large number of galaxies are very hard to be formed by chance.
Even when their mean densities are smaller than the average (i.e.,
$\kappa<0$), they are still very hard to be formed. Thus, when a
cluster containing the number of galaxies larger than 300 is
identified in an area as large as that of sample 1, it is unlikely
that this cluster is formed by chance.


\section{Identifying and classifying coherent clusters}

Applying our method to samples 1 and 3, we get entirely different
sets of coherent clusters. In sample 1, we detect 67 coherent
clusters under the condition that if any of them is formed by chance
then the probability will be less than that of $3\sigma $. Listed in
Table 2 are the parameters of these coherent clusters. Here, the
scale of a coherent cluster is defined as the largest value of the
distance measured for each pair of the sources of the cluster. The
smallest neighborhood function is $ \kappa =1.79$. For other 66
clusters, $\kappa >2$. The largest number of galaxies is 12966 and
the largest scale is $357h^{-1}Mpc$, which belong to different
coherent clusters. For all these clusters, $\kappa >1$, suggesting
that the number density, measured for each source within the scale
of $r_{ccs}(z)$, for any of these clusters is larger than twice of
the mean of the background sample. For sample 3, the largest number
of galaxies of the detected coherent clusters is 135 and the
neighborhood function of this cluster is $\kappa =1.22$. The number
of galaxies is larger than 134, but its neighborhood function is
slightly less than that coupling with 134, $\kappa =1.29$ (see
figure 12). Thus, the probability of forming this cluster by chance
is larger than that of $3\sigma $. The scattering distribution in
the plane of $(number,\kappa )$ of the coherent clusters detected
from sample 3 is shown in figure 13. Demonstrated in the figure,
there are no coherent clusters detected within the $3\sigma $
probability region for sample 3. Displayed in figure 13 is also the
scattering distribution of the coherent clusters detected from
sample 1. One finds from this figure that almost all of coherent
clusters detected from sample 3 are distributed beyond the $1\sigma
$ probability region in the plane of $ (number,\kappa )$. While a
sufficient number (67) of coherent clusters sorted out from sample 1
are situated within the $3\sigma $ probability region, there are
some others detected from the sample existing within the region
confined by the $1\sigma $ and $3\sigma $ probability contours.

\begin{figure}[tbp]
\begin{center}
\includegraphics[bb=0 0 460 460]{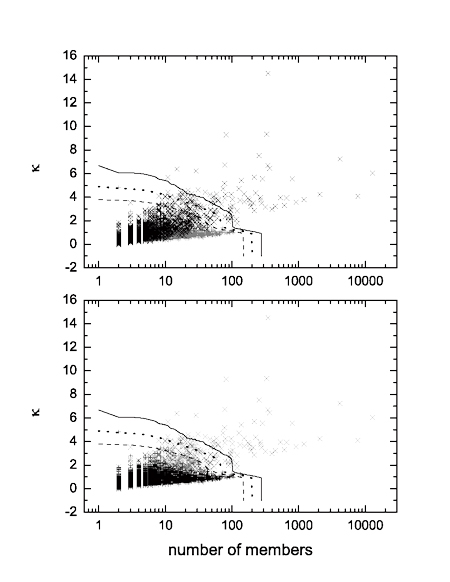}
\end{center}
\caption{Plot of $\kappa$ vs. number of galaxy members of coherent
clusters identified in samples 1 (crosses) and 3 (pluses), where the
probability contours in figure 12 are also presented.} \label{Figure
13}
\end{figure}

We define coherent clusters with probabilities being less than that of $%
3\sigma $ as prominent clusters, and define those with probabilities
being less than that of $1\sigma $ but larger than that of $3\sigma
$ as less prominent clusters, and call those with probabilities
being larger than that of $1\sigma $ as weak clusters.

\begin{figure}[tbp]
\begin{center}
\includegraphics[bb=0 0 460 560]{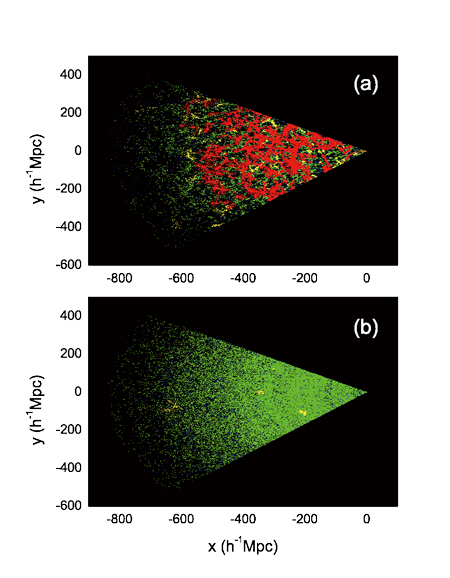}
\end{center}
\caption{Spatial (two dimensional) distributions of clustering
sources of samples 1 (panel a) and 3 (panel b). The blue color
stands for un-clustering sources, the green color represents sources
of weak clusters, the yellow color denotes sources of less prominent
clusters, and the red color symbolizes galaxies of prominent
clusters. For the definition of the coordinates, see figure 6.}
\label{Figure 14}
\end{figure}

\begin{figure}[tbp]
\begin{center}
\includegraphics[bb=0 0 460 560]{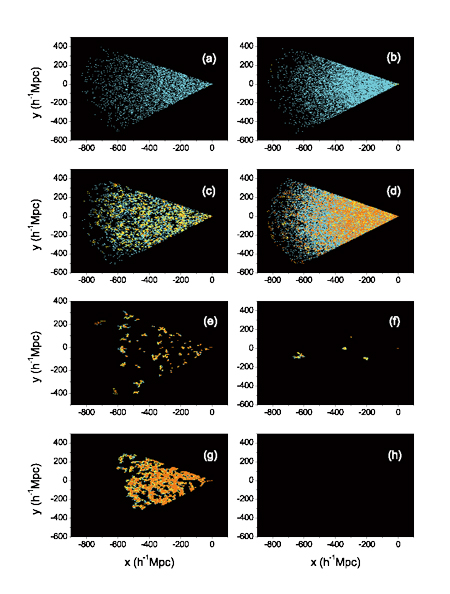}
\end{center}
\caption{Spatial (two dimensional) distributions of un-clustering
sources (panels a and b; the blue color in figure 14), weak clusters
(panels c and d; the green color in figure 14), less prominent
clusters (panels e and f; the yellow color in figure 14), and
prominent clusters (panels g and h; the red color in figure 14) for
samples 1 (panels a, c, e, and g) and 3 (panels b, d, f, and h). The
color symbols represent the same as they do in figure 6. For the
definition of the coordinates, see figure 6.} \label{Figure 15}
\end{figure}

Spatial (two dimensional) distributions of the three kinds of
cluster as well as the unclustered sources for the two samples are
shown in figures 14 and 15. We find that the unclustered sources of
both the observed and background samples are homogenously
distributed over the whole space concerned (see also figure 16
panels a and b). They act like what the low relative density ($
-1\leq \kappa _i<1$) galaxies do (see figure 9 panel a), and indeed,
almost all of them are low relative density galaxies (see figure 15
panels a and b). As suggested in panels (a) and (b) of figure 15,
the number of unclustered sources detected from sample 3 is larger
than that of sample 1. This is expectable since many original
unclustered regions (the presumed galaxies) might be pulled towards
a nearby relatively dense region (a presumed cluster) in later times
due to the continuous action of gravity. Figure 15 shows that the
majority of mid and high relative density sources (those of $ 1\leq
\kappa _i<3$ or $\kappa _i\geq 3$) are members of clusters (weak,
less prominent, or prominent clusters). Compared with the
corresponding panels in figure 16 we find that sources of weak
clusters in the background sample are homogenously distributed over
the whole volume of the sample as well, and the spatial distribution
of sources of this kind in the observational sample is a
quasi-homogenous one (see panels c and d in figures 15 and 16). In
addition, we find that the number of weak clusters in the background
sample is also larger than that in the observational sample. This
indicates that many unclustered galaxies and weak clusters become
members of large-scale structures formed in later times. As
expected, shown in panels (e), (f), and (g) of figure 15, we find
that the core of prominent and less prominent clusters is filled
mainly by sources with high number density neighborhoods ($3\leq
\kappa _i$), and surrounding them, galaxies with lower number
density environments are observed.

\begin{figure}[tbp]
\begin{center}
\includegraphics[bb=0 0 460 560]{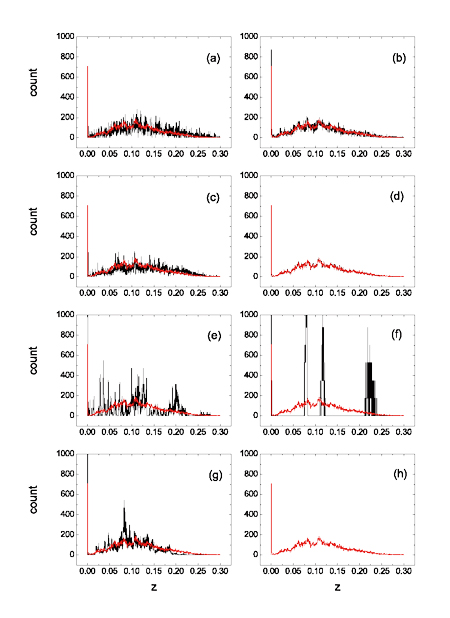}
\end{center}
\caption{Redshift distributions of un-clustering sources (black
lines in panels a and b), weak cluster sources (black lines in
panels c and d), less prominent cluster sources (black lines in
panels e and f), and prominent cluster sources (black lines in
panels g and h) for samples 1 (panels a, c, e, and g) and 3 (panels
b, d, f, and h). Red lines denote the redshift distribution of all
sources in sample 3 (black lines in figure 3 panel b). Numbers of
galaxies of all data sets are normalized to that of sample 1.}
\label{Figure 14}
\end{figure}

The numbers of unclustered, weak clustering, less prominent
clustering, and prominent clustering sources of sample 3 are 9033,
50126, 338 and 0, respectively. They occupy 15.2\%, 84.2\%, 0.568\%
and 0\% of the total number of the sample. The numbers of the
corresponding sources in sample 1 are 3557, 12700, 3017 and 40223
respectively, and they are 5.98\%, 21.3\%, 5.07\% and 67.6\% of the
total. This reveals that, in terms of statistics (i.e., under the
$3\sigma $ confidence level), the majority of galaxies observed in
the 2dF survey are members of prominent clusters (in about 67.6\%).
This indicates that the phenomenon of clustering is prominent in the
present Universe. Only a few galaxies (about 5.98\%) are identified
to be sources lonely staying in the space. Comparing the
corresponding numbers in the two samples we find that about 2/3
original unclustered sources and 3/4 original weak clustering
sources would change their identification in late times. It
indicates that original weak clustering sources have more chance to
become members of less prominent and prominent clusters in later
times and indeed they provide about 87\% of members of the latter
(note that only about 13\% members of the latter come from the
original unclustered sources). This is not surprising since the
seeds of clusters formed in later times are expected to be among
those originally resided in regions with higher number densities
(see panel d of figure 15). Displayed in figure 16 are redshift
distributions of unclustered, weak clustering, less prominent
clustering, and prominent clustering sources of samples 1 and 3. As
expected, redshift distributions of different kinds of source in the
background sample are well in agreement with the distribution of the
sample itself since no dynamics are at work in the stage of random.
(Note that the redshift distribution of less prominent clustering
sources is not in agreement with that of the whole sample, which is
due to the very small number of this kind of source, 0.568\%, of the
sample). While the redshift distribution of the unclustered sources
of the observational sample is consistent with that of the
background sample (see panel a of figure 16), there is a slight
deviation from that of the weak clustering sources of sample 1 to
that of the background sample, and an obvious deviation from that of
the prominent clustering sources of sample 1 to that of the latter
sample is observed. In addition we find that, for the observational
sample, there is a change from weak clusters to prominent clusters:
the number of sources in between (sources of less prominent
clusters) is relatively small.

According to this analysis, we suspect that there might exist three
kinds of galaxy or galaxy group in terms of clustering. One is the
class made up of isolated galaxies which would be homogenously
distributed over the Universe and would never be included in any
clusters. Positions of these galaxies would stick to the framework
of the Universe. Another class includes trivial coherent clusters
which would act like the isolated galaxies but their number
densities would evolve with time. The other class contains prominent
coherent clusters whose densities and numbers of galaxies would glow
all the time from the very early epoch to present. After the
clustering process of all galaxies in the Universe ceases, the
redshift distribution of the sources of these clusters would be that
of the homogenous spatial distribution, when measured in the volume
much larger than that associated with their typical scales. Measured
within the volume comparable to that associated with their typical
scales, it would not be surprising if the redhsift distribution
shows a character of in-homogenous spatial distribution. Before the
clustering process stoping, some potential members of these clusters
look like trivial coherent clusters in densities (described by
$\kappa$) and numbers of galaxies, and this would make the redshift
distribution of apparent trivial coherent clusters (weak clusters
defined in this paper; see panel c in figure 15) deviates from that
of the homogenously spatial distribution (see panel c in figure 16).
In this scheme, less prominent clusters defined in this paper should
be the potential members of the third class, and they would change
their identification in late times. According to this
interpretation, once a prominent cluster is formed it will maintain
its identification in late times. This explains why there is a turn
over change from the number of weak clustering sources to that of
prominent clusters observed above.

Revealed in figure 13, there are two approaches for a weak cluster
becoming a member of prominent clusters: increasing either its
density (described by $\kappa$) or its number of galaxy members. In
the first approach, they can accomplish the identification changing
themselves when their members become closer, while in the second
approach, the change would happen in the situation that they are
coherently connected with other clusters. For an isolated cluster,
when its relative density stops glowing, the process of
identification changing will be ended. This is why we expect a
homogenous spatial distribution of members of the second class.

The reasons that we propose the above scheme to interpret the
clustering process of galaxies are: a) we find it natural to explain
what we observe in the above clustering plots; b) it looks simple;
c) it can provide unambiguous predictions in the clustering process
which can be checked later; d) we find no simpler schemes accounting
for our results.

In the above analysis, we first classify four kinds of source
according to their clustering properties available from observation:
unclustered sources (which do not appear in figure 13), weak
clustering sources (which are beneath the $1\sigma $ contour in
figure 13), less prominent clustering sources (which are within the
area confined by the $1\sigma $ and $3\sigma $ contours in figure
13), and prominent clustering sources (which are above the $3\sigma
$ contour in figure 13). Later we propose a scheme with three kinds
of galaxy or galaxy group (isolated galaxies, trivial coherent
clusters, and prominent coherent clusters) to explain what we
observed in the statistical analysis. The four kinds of source might
change their identifications during the clustering process. But when
the clustering process ceases, each of them belongs to one of the
three kinds of galaxy or galaxy group.

\section{Structure of large prominent coherent clusters}

Here we show the structure of the two largest prominent coherent
clusters identified in sample 1. According to Table 2, the coherent
cluster with the largest number of galaxies is cluster 1 which
contains 12966 galaxies and extends to $ 281h^{-1}Mpc$; the one with
the second largest number of galaxies is cluster 2 which contains
less galaxy members (7788) but has the largest scale,
$357h^{-1}Mpc$. The scale of cluster 1 is the distance between the
two sources TGN141Z157 and TGN353Z218, and that of cluster 2 is the
distance between TGN163Z153 and TGN276Z102. Displayed in figure 17
are the two-dimensional positions (relative to that of other
galaxies of sample 1) of the two coherent clusters. Shown in figures
18 and 19 are their plane and solid structures, viewed from various
angles in the latter case.

\begin{figure}[tbp]
\begin{center}
\includegraphics[bb=0 0 560 290]{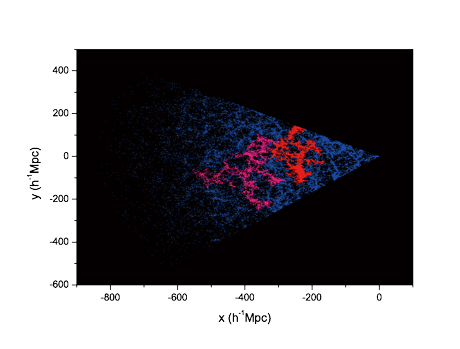}
\end{center}
\caption{Two-dimensional positions of the two largest coherent
clusters identified from sample 1, where the red and pink colors
represent clusters 1 and 2, respectively. Sources other than the two
clusters are presented in the blue color. For the definition of the
coordinates, see figure 6.} \label{Figure 17}
\end{figure}

\begin{figure}[tbp]
\begin{center}
\includegraphics[bb=0 0 560 540]{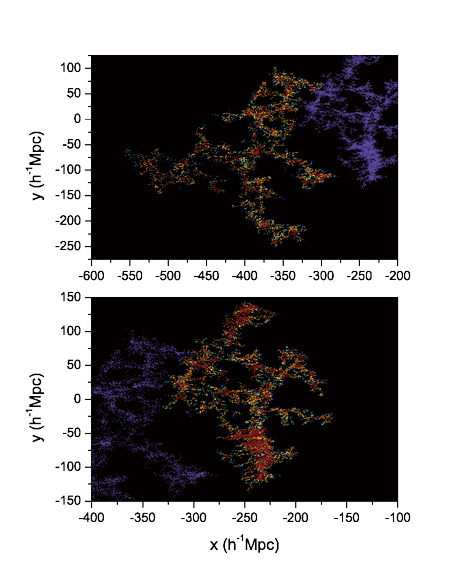}
\end{center}
\caption{Two-dimensional map of the structures of the two largest
coherent clusters detected from sample 1. Presented in the lower
panel is that of cluster 1, where the violet color represents the
sources of cluster 2. Shown in the upper panel is that of cluster 2,
where the violet color represents the sources of cluster 1. The
royal and blue colors stand for un-crowded sources (the royal color
for $-1 \leq \kappa _{i} < 0$ and the blue color for $0 \leq \kappa
_{i} < 1$), the cyan and yellow colors represent crowded sources
(the cyan color for $1 \leq \kappa _{i} < 2$ and the yellow color
for $2 \leq \kappa _{i} < 3$), and the orange and wine colors
symbolize very crowded sources (the orange color for $3 \leq \kappa
_{i} <6$ and the wine color for $6 \leq \kappa _{i}$). (Note that,
here we use six colors instead of only three to code different
crowded sources in order to show their spatial distributions in a
more detail. The colors used are somewhat different from those
adopted in figure 6.) For the definition of the coordinates, see
figure 6.} \label{Figure 18}
\end{figure}

\begin{figure}[tbp]
\label{Fig: 17}\centering
\includegraphics[bb=0 0 1200 580]{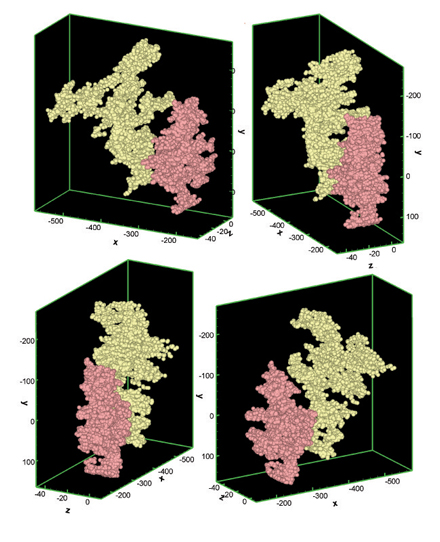}
\caption{Three dimensional map of the structures of the two largest
coherent clusters, where the yellow color represents cluster 1 and
the magenta color denotes cluster 2. Here, a relatively lower
resolution map is presented so that the solid structure is
noticeable. For the definition of the coordinates, see figure 6.}
\label{Figure 19}
\end{figure}

As shown in figure 17, the two clusters are quite close. The scales
of the two coherent clusters are comparable with the whole volume of
sample 1 (the upper limit of redshift in sample 1 is $z=0.3$ which
corresponds to $r\sim 837h^{-1}Mpc$; see also figure 17). This might
interpret why there is a bias of the redshift distribution of
prominent coherent clusters of sample 1 to the redhsift distribution
of the whole background sample (see figure 16 panel g, especially at
$z\sim 0.082$ which corresponds to $r\sim 241h^{-1}Mpc$). Revealed
by figure 18, very crowded sources form the core as well as the
frame of the corresponding structures. The two clusters possess
coral type structures in a fine map where the details of the
structure are visible. It seems that larger coherent clusters are
made up of smaller ones and in forming the former the latter are
likely connected by their antennae. This is in agreement with what
was discovered recently by Einasto et al. (2006b): superclusters are
asymmetrical and have multi-branching filamentary structure.

\begin{figure}[tbp]
\begin{center}
\includegraphics[bb=0 0 560 560]{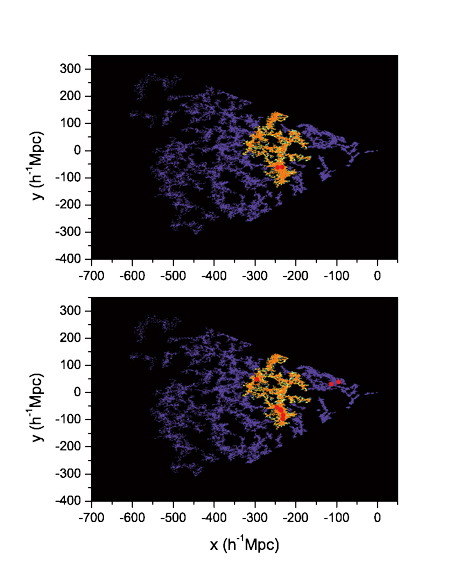}
\end{center}
\caption{Two-dimensional positions of clusters (the red color)
detected by Erdogdu et al. (2004) (the upper panel) and Eke et al.
(2004) (the lower panel). For the sake of comparison, cluster 1
identified in this paper from sample 1 is presented with the same
coded colors adopted in figure 6. Sources of other prominent
clusters identified from sample 1 are also presented which are coded
with the violet color. The two filled circles in the upper panel
denotes the the redshift range where SCNGP06 (Erdogdu et al. 2004)
is detected and located. The filled circles in the lower panel stand
for central positions of the nine large galaxy number groups
detected from NGP, presented in Table A2 of Eke et al. (2004). For
the definition of the coordinates, see figure 6.} \label{Figure 20}
\end{figure}

In the same volume studied above, Erdogdu et al. (2004) detected a
large cluster, SCNGP06. We find that SCNGP06 is just inside the
structure of cluster 1, which is shown in figure 20. Eke et al.
(2004) also identified large galaxy number groups in the same
volume. Among their 9 largest groups identified within the volume, 7
are inside the structure of cluster 1 (see figure 20). Revealed in
the figure, centers of SCNGP06 and the 7 large galaxy number groups
of Eke et al. (2004) are located at the knots of the structure.

\begin{figure}[tbp]
\begin{center}
\includegraphics[bb=0 0 460 560]{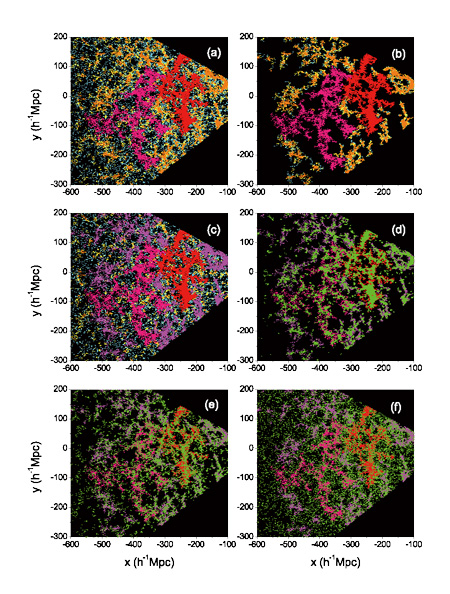}
\end{center}
\caption{Two-dimensional structures of clusters 1 (red) and 2 (pink)
and the spatial distribution of different crowded galaxies of sample
1 within the same area. In panel (a), colors other than the red and
pink denote all sources other than those of the two clusters within
the area, and these sources are coded with the same colors as they
are coded in figure 6. Panel (b) is a copy of panel (a), where
sources other than those of prominent clusters are taken off. Panel
(c) is also a copy of panel (a), where sources of prominent clusters
other than clusters 1 and 2 are symbolized by the magenta color. In
panels (d), (e), and (f), the red, pink, and magenta colors
represent sources of cluster 1, cluster 2, and all other prominent
clusters, respectively. Green colors in panels (d), (e), and (f)
stand for all very crowded, crowded, and un-crowded sources in
sample 1 within the area, respectively. For the definition of the
coordinates, see figure 6.} \label{Figure 21}
\end{figure}

As suggested above (see figure 9 panels b and c), voids are likely
the volumes within which no crowded sources are present. In the
structures of the two largest coherent clusters, voids are easily
identified (see figure 18). Are they really the space where no
crowded sources are found? Figure 21 shows the structures of the two
clusters together with that of other prominent clusters and the
spatial distribution of different crowded galaxies of sample 1
within a smaller area enclosing the structures. We find from panel
(a) of the figure that, for the majority of voids detected from the
structures of the two clusters, there indeed no or very few high
density galaxies are present. This can be clearly seen in panels (d)
and (e). Within the apparent void of $(x,y)\sim (-300,-50)h^{-1}Mpc$
in panel (a), there are some very crowded sources. We suspect that
this might not be a real void, but instead, it might come from the
projected effect. In other words, the inner structures observed
within the apparent void might not be close to the two clusters in
the three-dimensional space, but they look like to be close to them
in the projected two-dimensional space. Indeed, as revealed in
panels (b) and (c), inside the apparent void, some structures of
prominent clusters are present. Shown in panel (d), there are indeed
few very crowded sources outside the structures of prominent
clusters, and it is these galaxies that form the core and build the
frame, of the structures. Crowded sources tend to be gathered around
the structures, as seen in panel (e). Just as pointed out above,
un-crowded sources are seen to be homogenously distributed over the
structure area, and they are indeed detected within the voids as
well as within the frame of the structures (see panel f). In a
recent investigation of voids, Patiri et al. (2006) showed that
faint galaxies populating the voids are clustered in small groups
and filaments. This is observed in panel (f). In addition, we find
that the apparent filaments of un-crowded galaxies do not stretch
along the frame of the structure of the prominent clusters, as what
the very crowded and crowded sources do in panels (d) and (e).
Recalled that filamentary features could appear in the state of
random distribution (see the discussion in Section 4), we thus
suspect that the apparent filaments of low density galaxies might
mainly be caused by fluctuation.

\section{The largest structure detected in a larger data set of the 2dF
survey}

One might observe that, in the method adopted above, while the
calculation of the relevant probability and the investigation of
number density distribution depend on the exact volume of the
adopted sample, the sorting out method (i.e., the FoF algorithm) is
independent of the volume. This enables us to identify the largest
scale structure of a sample of the same survey, for which, the
borders are not well defined and the volume is not available.
Although the probability of forming such structure by chance is
unable to be evaluated, at least the following questions might be
able to answer. How large is the scale of the largest structure
identified from the whole 2dF survey by the FoF algorithm? Could
some unconnected structures identified in a smaller volume be
coherently connected in a larger volume, which is strongly expected?
Is the coral type structure detected in the space of sample 1 merely
a local nature? (Or, could it be found in other areas of the
survey?).

Here, we apply the sorting out method proposed above to the whole
2dFGRS data set. To match the basic criterions of sample 1, we
simply adopt sample 2 for our analysis. The $ \rho_{0}(z)$ used to
determine the $r_{ccs}$ is that directly measured from sample 2
(i.e., $\rho1$ in Table 1).

\begin{figure}[tbp]
\begin{center}
\includegraphics[bb=0 0 760 280]{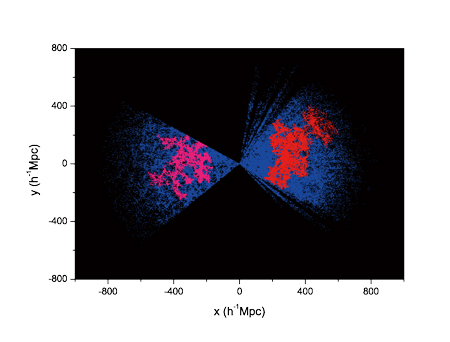}
\end{center}
\caption{Two-dimensional positions of the two largest coherent
clusters detected from sample 2, where the red and pink colors
represent the largest and the second largest coherent clusters,
respectively. Sources other than the two clusters are presented in
the blue color. For the definition of the coordinates, see figure
6.} \label{Figure 19}
\end{figure}

\begin{figure}[tbp]
\begin{center}
\includegraphics[bb=0 0 560 580]{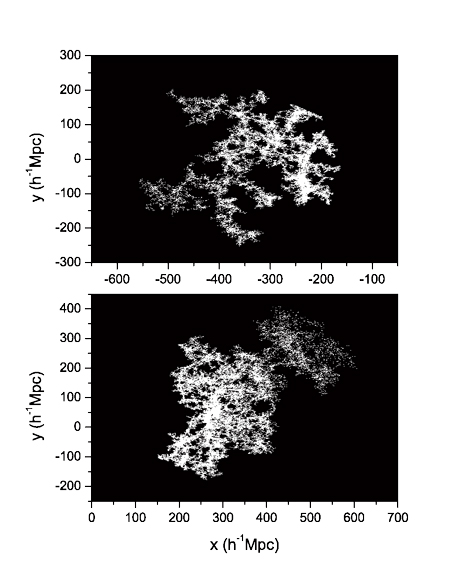}
\end{center}
\caption{Two-dimensional map of the structures of the two largest
coherent clusters detected from sample 2. Presented in the lower
panel is that of the largest coherent cluster, and shown in the
upper panel is that of the second largest coherent cluster. For the
definition of the coordinates, see figure 6.} \label{Figure 20}
\end{figure}

\begin{figure}[tbp]
\label{Fig: 21}\centering
\includegraphics[bb=0 0 1050 480]{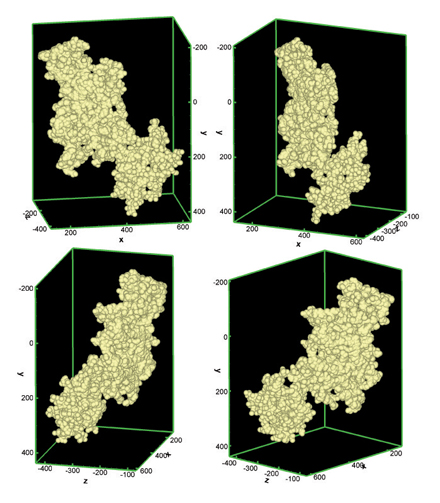}
\caption{Three dimensional map of the structure of the largest
coherent cluster identified from sample 2. Here, a relatively lower
resolution map is presented so that the solid structure is
noticeable. For the definition of the coordinates, see figure 6.}
\end{figure}

The largest number of galaxies of the coherent clusters identified
in sample 2 is 41813, and the scale of this cluster is
$659h^{-1}Mpc$ which is the largest one detected with the adopted
method. The second largest cluster contains 32188 galaxies and its
scale is $474h^{-1}Mpc$. The two-dimensional positions (relative to
that of other galaxies of sample 2) of the two coherent clusters are
illustrated in figure 22 which shows that the structures of the two
groups of galaxies are situated in two distinct areas of the sky.
Comparing it with figure 17 we find that the two largest coherent
clusters identified from sample 1, clusters 1 and 2, now become a
single one, the second largest cluster identified from sample 2 (the
one shown in the left-hand side of figure 22). Besides those of
clusters 1 and 2, some other sources are also included in the second
largest cluster. This is expectable, since the area of sample 1 is
well inside that of sample 2 in that direction of the sky (see
figure 1), and some bridge sources outside the area of sample 1
might exist and then would connect clusters 1 and 2 together as well
as connect other sources or clusters to them. The right-hand side of
figure 22 shows that large-scale structures could extend to the
space as far as the adopted samples could reach (the largest
redshift detected in the right-hand side structure, the largest
coherent cluster detected from sample 2, is $z=0.2646$). (For the
left-hand side structure, the largest redshift detected is
$z=0.1969$.) This indicates that the large-scale structure is not a
patent right of nearby galaxies. Instead, it could be formed in a
quite early time, although space sampling is sparse at that epoch
(see figure 8 panel a). In fact, it was found in recent observations
that large structures are common at redshifts as large as $z\sim 3$
(Steidel et al. 1998). Large-scale structures which are made up of
protoclusters could be formed even at the epoch as early as $z\sim
6$ (Ouchi et al. 2005).

The structure of the largest cluster detected from sample 2, that in
the right-hand side of figure 22, is displayed in figures 23-24,
where more details are visible (shown in figure 23 is also the two
dimensional structure of the second largest cluster). Figure 23,
which is a fine map, shows that the topology of the structure is
coral-like as well.

Although the probability for forming the two largest coherent
clusters by chance is not available due to the ill-defined borders
of sample 2, we tend to believe that the two clusters are not formed
by chance. According to figure 12, large numbers of galaxies of
coherent clusters are scarcely formed by chance, even their mean
relative densities are small. In sample 3, the largest number of
galaxies is much less than 300 (see figure 13), but the largest one
detected from sample 1 is 12966 which is 43 times of 300. The
largest and the second largest numbers of galaxies of the coherent
clusters identified from sample 2 are 139 and 107 times of 300,
respectively. It is unlikely that the enhancement of the volume from
sample 1 to sample 2, where the number of galaxies of the latter is
about 4 times of the former, would shift the $3\sigma $ probability
contour in figure 13 to the number of galaxies as large as that of
the second largest cluster, 32188 (a detailed analysis on this issue
will be performed later).

\section{Discussion and conclusions}\label{s.conc}

Our analysis is performed in redshift space, where the peculiar
velocity of individual sources is ignored. As is known, in measuring
the distance of an object, a big problem is the joining of its
dynamical velocity to the redshift observed (see, e.g., Gellar \&
Huchra 1989; Guzzo 2002). Generally, the velocity distortion would
shrink overdense regions and inflate underdense ones (Kaiser 1987;
Martinez \& Saar 2002a). As shown in figure 13, correcting this
effect would raise some weak clusters of sample 1 to be less
prominent clusters since their number densities (represented by
$\kappa$) would become larger. In contrast, the correction would
make some prominent clusters to be less prominent ones since their
number densities would become smaller.

When the redshift distortion is removed, the pattern of clustering
observed above might be altered. However, as revealed by figure 18,
the possible change of the two largest structures of sample 1 would
not be so severe, since the structures themselves are firmly built
on the very dense frames.

The measurement uncertainty of redshift for the 2dF galaxy redshift
survey is $85 km s^{-1}$ (Colless et al. 2001), which corresponds to
$\Delta z = 2.8\times 10^{-4}$. The line of sight positions will
certainly be affected by errors in redshifts. Would this play an
important role in identifying clusters, where the FoF algorithm is
applied? The answer is yes when the redshift concerned is as small
as $z<0.001$. Accordingly, as revealed by Table 2, clusters 14, 16,
22, 33, 47, 52, 54, 61, and 64 will certainly be affected by the
redshift measurement uncertainty. When much smaller redshift
uncertainty is available, some of these clusters might be
disassembled while some of them might be merged into other clusters.
However, as illustrated in figure 3, most sources in sample 1 have
their redshifts much larger than the uncertainty and thus the effect
of redshift errors must be very limited in identifying clusters
among them. In particular, the largest clusters discussed above are
located in relatively high redshift regions (see figure 22 and Table
2). They would be much less affected by this uncertainty.

Colless et al. (2001) pointed out that the survey magnitude limit of
2dFGRS varies slightly with position on the sky. In the SGP the
median limiting magnitude is $b_{j}=19.40$, with an rms of 0.05 mag;
in the NGP the median limiting magnitude is $b_{j}=19.35$, with an
rms of 0.11 mag. To show how this effect plays a role, we repeat the
grouping analysis performed above by adopting a brighter magnitude
cut so that the sample selected will be more complete in magnitude.
According to Colless et al. (2001), we take the limiting magnitude
as $b_{j}=19.24$ and consider only galaxies brighter than it. Under
this condition, there are 198647 sources left in sample 2 ($87.8\%$)
and 55846 galaxies remained in sample 1 ($93.9\%$). Due to the cut
in the total number of the samples, a straightforward result is that
all clusters identified above have smaller members and smaller
values of $\kappa$. This is true. Cluster 1 now has only 12391
sources with $\kappa = 5.90$. For cluster 2, the number of members
now becomes 6237 and the value of $\kappa$ is 3.97. The scale of
cluster 1 remains unchanged, while that of cluster 2 now becomes
$303 h^{-1} Mpc$. This is not surprising since faint sources are
always far away sources and cluster 2 is identified in a larger
distance than that of cluster 1. Spatial distributions of the two
clusters are shown in Fig. 25, where, for the sake of comparison,
the original members of the two clusters are also presented. One
finds that while cluster 1 remains almost the same, cluster 2 is
obviously affected. Some original substructures of cluster 2 are
dismissed while some new substructures are attached. The following
previous conclusion holds: clusters could stretch to large
distances. (What would happen when continuously adopting brighter
magnitude limits? Currently we have no answers to this. It deserves
an intensive investigation.) Besides the number and density of
clusters, the new magnitude limit might also affect the possibility
contour of sample 1. Since the total number of the sample becomes
smaller than the previous one, clusters identified with the same
algorithm would have smaller numbers and would be less dense. In
this way, the contours shown in Figs. 12 and 13 would shift
leftwards. As revealed by Fig. 13, clusters 1 and 2 are far away
from the $3\sigma$ contour, even though they have smaller numbers
than before due to the brighter limit. The conclusion that the
probability of picking the two clusters by chance is much less than
that of $3\sigma$ maintains.

\begin{figure}[tbp]
\begin{center}
\includegraphics[bb=0 0 560 480]{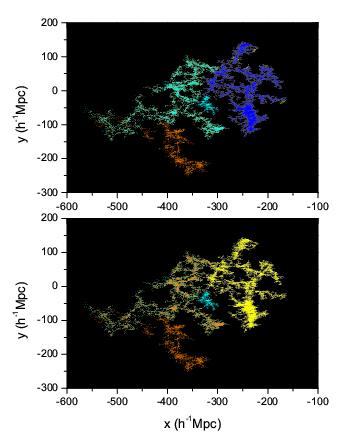}
\end{center}
\caption{Two-dimensional map of the structures of the two largest
coherent clusters detected from sample 1 when a brighter magnitude
limit is adopted. The blue color stands for cluster 1 and the cyan
color for cluster 2. The old galaxy members of the the two clusters
are also presented for the sake of comparison, where the yellow
color represents the old cluster 1 and the orange color denotes the
old cluster 2.} \label{Figure 25}
\end{figure}

Some authors proposed that the three dimensional topology of
large-scale structures is spongelike (Gott III et al. 1986; Vogeley
et al. 1994; Colley et al. 2000). Indeed, when plotting the
structures of the two largest clusters identified from sample 1 in a
lower resolution map, this type of feature emerges (see figure 25).
It suggests that the large-scale structure of the local Universe is
intrinsically coral-like, but it could also be spongelike if the
resolution of the map is low enough.

\begin{figure}[tbp]
\begin{center}
\includegraphics[bb=0 0 460 300]{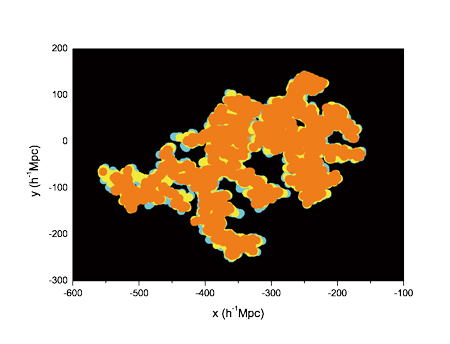}
\end{center}
\caption{Low resolution (two-dimensional) map of the structures of
the two largest coherent clusters identified from sample 1. The
colors denoting different crowded sources are the same as they are
coded in figure 6. For the definition of the coordinates, see figure
6.} \label{Figure 22}
\end{figure}

As already pointed out by Gott III et al. (2005) that some larger
structures are found if larger samples are involved. This is natural
and is indeed observed in the analysis on sample 2, where the two
large-scale structures detected in a smaller sample (sample 1) merge
and a much larger structure is identified (see figure 22).
(Accordingly, as the volume of sample 2 occupies only a small
fraction of the nearby space, it is suspected that we might be able
to observe a structure with its scale as large as $\sim
1000h^{-1}Mpc$ in the local Universe so long as the whole sky is
surveyed.)

Implied by figure 21, some of the coherent clusters and unclustered
sources seem to be able to connect with each other in a
two-dimensional view. This is indeed true. When one repeats the
above analysis on sample 1 under the condition that the contribution
of the vertical length (i.e., the contribution of the coordinate $z$
defined in figure 6) is ignored, one identifies a much larger
coherent cluster than the largest one obtained above (the details of
the analysis are omitted). This suggests that some large-scale
structures observed in a two-dimensional map might not be a coherent
one when the vertical length is taken into account. As shown in
Table 2, the mean redshift of cluster 1 is 0.0864. At this redshift,
the vertical length confined by the adopted window of sample 1 (say,
$ -4.8^{\circ }<Dec.(J2000)<1.0^{\circ }$) is $25.7h^{-1}Mpc$, while
the adopted criterion clustering scale at this redshift is
$r_{ccs}=3.07h^{-1}Mpc$. The former is about 8 times of the latter.
Keeping this in mind, the phenomenon that some structures connected
in a two-dimensional map are in fact separated is expectable.


Note that $r_{ccs}$ might be different for different surveys. It is
expected that the number of galaxies we will eventually ever be able
to see would be larger than that observable today by a factor of
$2.36$ (Gott III et al. 2005). If sample 1 is enlarged by two times,
$r_{ccs}$ might probably shrink to a smaller scale. It is
unnecessary that this would lead to a smaller scale of the largest
structure identified with $r_{ccs}$. One reason is that the
enlarging of the sample makes the sources more crowded and then
separated sources are easier to be connected. Another reason is that
the largest structure is constructed with the frames made up of very
crowded sources (with several times of number density than that of
the background data). The enlargement of the sample (only two times
larger) would not be able to change this situation.

It should be pointed out that the structure scale as large as
$357h^{-1}Mpc$ obtained from sample 1 and that as large as
$659h^{-1}Mpc$ gained from sample 2 depend on the criterion
clustering scale adopted in this paper, $ r_{ccs}=[2\rho
_0(z)]^{-1/3}$. It is obvious that when a smaller $r_{ccs}$ is
adopted, then a smaller value of the largest structure scale
identified from the samples would be obtained, and when a larger one
is taken, then one would get a larger value of the largest structure
scale. Thus, talking the largest structure scale one should refer to
the corresponding criterion clustering scale. Due to this and due to
the fact that the criterion clustering scale adopted in this paper
varies with redshift, the comparison of the largest structure scale
between our analysis and other previous ones has less meaning. It is
desired that we can construct a standard criterion clustering scale
one day and then can compare the largest structure scale obtained
from various samples or from different regions of the Universe.
Although we are unable to compare our results with others, the
algorithm adopted in this paper tends to pick out larger structures
than what identified previously, when similar criterion clustering
scales are adopted. It is because that we are identifying
large-scale structures in the co-moving frame of the Universe rather
than in the observer frame. It is natural that those galaxies
excluded in the previously identified large-scale structures due to
their large redshifts are possible becoming members of some
large-scale structures identified in this paper when they are
located in a relatively dense region (see Fig. 22).

Martinez et al. (1998) searched for the scale of homogeneity by
applying the $K$ function to galaxy catalogues. They detected the
fingerprint of the transition to homogeneity in all the cases
considered. Amendola \& Palladino (1999) analyzed volume-limited
subsamples of a redshift survey to search for the scale of
homogeneity and found that the survey shows a trend of homogeneity
at large scales. As shown in figure 4, one finds a trend toward
homogeneity as well. We argue that, at least what revealed by figure
4 could not be interpreted as the existence of a transition to
homogeneity at a large scale since that scale is comparable to the
maximum scale of the adopted sample (see the caption of figure 4).
As pointed out in Section 3, we suspect that the flux limit as well
as the possible evolutionary effect might be the main factors that
block the detection of the real scale of homogeneity. The previous
methods cannot solve this problem, neither can the neighborhood
function introduced in this paper. Only when this effect is removed,
one will be able to determined the homogeneity scale with galaxy
redshift surveys.

It may be an intrinsic property that this method naturally produces
larger structures than other approaches do.

As suggested above, filamentary features are observed in the state
of random distribution, formed by crowded sources (see figure 10).
This is in agreement with the previous prediction that galaxies
preferentially formed in large-scale filamentary or sheet-like mass
overdensities in the early Universe, which was detected at as far as
$z=3.1$ (Matsuda et al. 2005). When sorting out all sources of the
background sample with $r_{ccs}$ we get many weak clusters and few
less prominent clusters (see figure 15 panels d and f). According to
definition, these clusters are likely the so-called protoclusters
(Venemans 2005). The latter were identified at $z\sim 4$ in recent
observations (Venemans et al. 2002; Miley et al. 2004; Intema et al.
2006).

According to the above analysis we conclude that: 1) the probability
of forming the large-scale structures, detected from sample 1 with
our method, by chance is very small; 2) the phenomenon of clustering
is dominant in the local Universe; 3) coherent clusters with the
scale as large as $357h^{-1}Mpc $ and the number of galaxies as
large as 12966 are identified from sample 1 well within the $3\sigma
$ confidence level; 4) there exist some galaxies which are not
affected by the gravitation of clusters and therefore are likely to
rest on the co-moving frame of the Universe; 5) filamentary features
could appear in the state of random distribution; 6) voids are
likely the volumes within which no very crowded sources are present
and they are likely formed in embryo by fluctuation in the very
early epoch of the Universe and it might be the continuous
gravitation in later time that pulls crowded galaxies closer and in
the same time leaves behind adult voids; 7) large-scale structures
are coral-like and they are likely made up of smaller ones and in
forming the former the latter are likely connected by their
antennae; 8) very crowded sources are mainly distributed within the
structure of prominent clusters and it is them who form the frame of
the large-scale structure.



\section*{Acknowledgements}

This work was supported by the National Science Fund for
Distinguished Young Scholars (10125313), the National Natural
Science Foundation of China (No. 10573005), and the Fund for Top
Scholars of Guangdong Provience (Q02114). We also thank the
financial support from the Guangzhou Education Bureau and Guangzhou
Science and Technology Bureau.

\clearpage

\begin{table}[tbp]
\caption{Description of samples.}
\renewcommand{\baselinestretch}{0.85}
\small
\begin{center}
\begin{tabular}{cccccc}
\\ \hline  \hline

name & number & area & region$^{a}$ & $ \rho_{0}(z)$$^{c}$ & data \\
\hline

sample 1 & 59497 & $150.0^{0}<RA<220.0^{0}$ & NGP & $ \rho1$ & observation \\
         &       & $-4.8^{0}<DEC<1.0^{0}$ & & & \\
sample 2 & 226302 & The whole area of the 2dF survey & NGP, SGP$^{b}$ & $ \rho1$ & observation \\
sample 3 & 59497 & That of sample 1 & NGP & $ \rho1$ & simulation \\
sample 4 & 59497 & That of sample 1 & NGP & $ \rho2$ & simulation \\
sample 5 & 59497 & That of sample 1 & NGP & $ \rho3$ & simulation \\

\hline
\end{tabular}
\end{center}

a: For definitions of NGP and SGP, see Colless et al. (2001).\\
b: In addition to NGP and SGP, sample 2 covers all other `random'
fields of the 2dF survey. \\
c: In this column we list the $ \rho_{0}(z)$ that is used to perform
the corresponding simulations, to calculate the corresponding
$\kappa$, and/or to determine the corresponding $r_{ccs}$, where $
\rho1$ is that directly measured from sample 2, $ \rho2$ is the
polynomial function obtained by a fit to $ \rho1$, and $\rho3=\rho
_0=0.00247(Mpc/h)^{-3}$ is a constant.
\end{table}

\clearpage

\begin{table}[tbp]
\caption{Parameters of the $67$ coherent clusters with $P<P(3\sigma
)$ detected from sample 1.}
\begin{center}
\begin{tabular}{ccccc|ccccc}\\ \hline\hline
label & number & $\kappa(r_{ccs})$ & scale$^a$ & $\overline{z}$ &
label & number & $\kappa(r_{ccs})$ & scale & $\overline{z}$ \\
\hline

1 & 12966 & 6.04 & 281 & 0.0864 & 35 & 128 & 4.73 & 12.0 & 0.0238 \\
2 & 7788 & 4.07 & 357 & 0.136 & 36 & 121 & 2.82 & 34.2 & 0.136 \\
3 & 4216 & 7.24 & 136 & 0.0485 & 37 & 121 & 1.79 & 106 & 0.218 \\
4 & 2100 & 4.24 & 177 & 0.130 & 38 & 118 & 2.11 & 71.5 & 0.201 \\
5 & 1215 & 3.78 & 112 & 0.124 & 39 & 103 & 3.27 & 56.2 & 0.184 \\
6 & 922 & 3.56 & 127 & 0.173 & 40 & 102 & 3.82 & 58.4 & 0.185 \\
7 & 805 & 5.56 & 51.9 & 0.0533 & 41 & 98 & 4.22 & 19.5 & 0.0790 \\
8 & 668 & 3.93 & 62.5 & 0.111 & 42 & 97 & 2.84 & 29.6 & 0.105 \\
9 & 615 & 3.29 & 121 & 0.182 & 43 & 95 & 2.92 & 26.4 & 0.0958 \\
10 & 540 & 3.96 & 56.6 & 0.0748 & 44 & 95 & 3.18 & 26.4 & 0.0829 \\
11 & 448 & 3.22 & 46.7 & 0.0821 & 45 & 82 & 9.29 & 13.1 & 0.0086 \\
12 & 401 & 2.97 & 125 & 0.167 & 46 & 81 & 4.24 & 15.9 & 0.0175 \\
13 & 369 & 6.45 & 28.7 & 0.0264 & 47 & 79 & 6.73 & 0.714 & 0.0005 \\
14 & 349 & 14.5 & 0.343 & 0.0001 & 48 & 79 & 3.88 & 28.8 & 0.108 \\
15 & 348 & 6.70 & 26.3 & 0.0236 & 49 & 78 & 5.17 & 16.8 & 0.0759 \\
16 & 333 & 9.34 & 0.688 & 0.0002 & 50 & 74 & 4.73 & 22.8 & 0.100 \\
17 & 321 & 3.16 & 70.6 & 0.166 & 51 & 70 & 3.39 & 58.2 & 0.200 \\
18 & 287 & 5.56 & 26.1 & 0.0476 & 52 & 63 & 3.72 & 0.512 & 0.0004 \\
19 & 280 & 3.48 & 68.2 & 0.105 & 53 & 62 & 3.78 & 12.4 & 0.0304 \\
20 & 271 & 3.74 & 89.8 & 0.189 & 54 & 59 & 4.25 & 1.09 & 0.0006 \\
21 & 269 & 3.95 & 59.9 & 0.103 & 55 & 56 & 3.78 & 25.9 & 0.107 \\
22 & 257 & 7.52 & 0.900 & 0.0003 & 56 & 55 & 3.46 & 23.7 & 0.115 \\
23 & 228 & 4.37 & 24.1 & 0.0203 & 57 & 53 & 4.08 & 8.54 & 0.0044 \\
24 & 218 & 5.70 & 18.7 & 0.0201 & 58 & 51 & 3.74 & 40.3 & 0.214 \\
25 & 216 & 2.86 & 92.1 & 0.150 & 59 & 47 & 4.23 & 12.4 & 0.0535 \\
26 & 211 & 3.51 & 64.6 & 0.145 & 60 & 40 & 4.42 & 13.6 & 0.100 \\
27 & 194 & 3.21 & 32.6 & 0.0475 & 61 & 31 & 4.61 & 0.477 & 0.0007 \\
28 & 188 & 3.46 & 56.6 & 0.144 & 62 & 29 & 6.21 & 8.96 & 0.103 \\
29 & 182 & 3.77 & 52.4 & 0.132 & 63 & 25 & 4.27 & 9.30 & 0.0654 \\
30 & 175 & 3.90 & 40.3 & 0.0689 & 64 & 22 & 4.59 & 0.320 & 0.0006 \\
31 & 155 & 3.58 & 55.1 & 0.165 & 65 & 22 & 4.31 & 10.7 & 0.141 \\
32 & 154 & 4.28 & 23.4 & 0.0648 & 66 & 15 & 6.39 & 7.51 & 0.147 \\
33 & 139 & 5.43 & 0.823 & 0.0004 & 67 & 11 & 5.69 & 7.47 & 0.125 \\
34 & 133 & 3.72 & 39.6 & 0.136 &  &  &  &  &  \\

\hline
\end{tabular}
\\[0pt]
\par
a: in units of $h^{-1}Mpc$.
\end{center}
\end{table}

\end{document}